\documentclass[11pt]{article}

\usepackage{amsmath}
\usepackage{rotating}
\usepackage{graphicx,subfigure}

\include{definitions}

\def \IR{\hbox{{\rm I}\kern-.2em\hbox{{\rm R}}}}

\newcommand{\bX}{\mbox{\boldmath $X$}}
\newcommand{\bY}{\mbox{\boldmath $Y$}}

\newcommand{\bx}{\mbox{\boldmath $x$}}

\newcommand{\N}{\mbox{$N$}}

\newcommand{\bbeta}{\mbox{\boldmath $\beta$}}

\newcommand{\bgamma}{\mbox{\boldmath $\gamma$}}

\usepackage[parfill]{parskip}
\usepackage{epsfig, subfigure, amsfonts}
\usepackage{amsfonts}
\usepackage{natbib}

\usepackage{url}
\usepackage{authblk}

\usepackage[margin=1in]{geometry}

\usepackage{booktabs}
\usepackage{bigstrut}
\usepackage{multirow}
\usepackage{threeparttable} 
\usepackage[format=hang,labelfont=bf]{caption}
\usepackage{colortbl}
\definecolor{light-gray}{gray}{0.75}
\usepackage{url}
\usepackage{hyperref}

\usepackage[toc,page]{appendix}

\usepackage[rightbars, color]{changebar}
\cbcolor{green}
\begin{document}

\pagenumbering{roman}

\title{{\bf {\sc Hyak} Mortality Monitoring System}\\Innovative Sampling and Estimation Methods\\{\large Proof of Concept by Simulation}}

\author[1,2,3,4,*]{Samuel J. Clark}
\author[5,6]{Jon Wakefield}
\author[5,7]{Tyler McCormick}
\author[8]{Michelle Ross}

\affil[1]{Department of Sociology, The Ohio State University}
\affil[2]{MRC/Wits Rural Public Health and Health Transitions Research Unit (Agincourt), \authorcr School of Public Health, Faculty of Health Sciences, University of the Witwatersrand}
\affil[3]{INDEPTH Network, Accra}
\affil[4]{ALPHA Network, London}
\affil[5]{Department of Statistics, University of Washington}
\affil[6]{Department of Biostatistics, University of Washington}
\affil[7]{Department of Sociology, University of Washington}
\affil[8]{Department of Biostatistics and Epidemiology, University of Pennsylvania}
\affil[*]{Correspondence to: \texttt{work@samclark.net}}

\date{\today}

\maketitle

\centerline{ {\sc Abstract} }
\small

Traditionally health statistics are derived from civil and/or vital registration.  Civil registration in low- to middle-income countries varies from partial coverage to essentially nothing at all.  Consequently the state of the art for public health information in low- to middle-income countries is efforts to combine or triangulate data from different sources to produce a more complete picture across both time and space -- {\em data amalgamation}. Data sources amenable to this approach include sample surveys, sample registration systems, health and demographic surveillance systems, administrative records, census records, health facility records and others. 

We propose a new statistical framework for gathering health and population data -- {\sc Hyak} -- that leverages the benefits of sampling and longitudinal, prospective surveillance to create a cheap, accurate, sustainable monitoring platform.  {\sc Hyak} has three fundamental components:

\begin{itemize}
\item {\bf Data Amalgamation}: a sampling and surveillance component that organizes two or more data collection systems to work together: (1) data from HDSS with frequent, intense, linked, prospective follow-up and (2) data from sample surveys conducted in large areas surrounding the Health and Demographic Surveillance System (HDSS) sites using informed sampling so as to capture as many events as possible;
\item {\bf Cause of Death}: verbal autopsy to characterize the distribution of deaths by cause at the population level; and 
\item {\bf SES}: measurement of socioeconomic status in order to characterize poverty and wealth.
\end{itemize}

We conduct a simulation study of the informed sampling component of {\sc Hyak} based on the Agincourt HDSS  site in South Africa.  Compared to traditional cluster sampling, {\sc Hyak}'s informed sampling captures more deaths, and when combined with an estimation model that includes spatial smoothing, produces estimates of both mortality counts and mortality rates that have lower variance and small bias.  


\normalsize

\newpage

\centerline{\MakeUppercase{Acknowledgments}}

Preparation of this manuscript was supported by the Bill and Melinda Gates Foundation and grants K01 HD057246 and K01 HD078452 from the Eunice Kennedy Shriver National Institute of Child Health and Human Development (NICHD). JW was supported by R01-CA095994. The authors are grateful to Peter Byass, Basia Zaba, Stephen Tollman, Adrian Raftery, Philip Setel, Osman Sankoh and two very constructive anonymous reviewers for helpful discussions or other inputs.

\newpage

\tableofcontents

\newpage

\pagenumbering{arabic}

\section{New Directions for Health and Population Statistics in Low- to Middle-Income Countries}


\subsection{Background}

In most of the developed world the traditional source of basic public health information is civil registration and vital statistics.  Civil registration is a system that records births and deaths within a government jurisdiction.  The purpose is two-fold: (1) to create a legal record for each person, and (2) to provide vital statistics.  Optimally a civil register includes everyone in the jurisdiction, provides the basis to ensure their civil rights and creates a steady stream of vital statistics \citep{united2014principles}.

The vital statistics obtained from many well-functioning civil registration systems include birth rates by age of mother, mortality rates by sex, age and other characteristics, and causes of death for each death.  These basic indicators are the foundation of public health information systems, and when they are taken from a near-full-coverage civil registration system, they relate to the whole population.

Although the idea is inherently simple, implementing full-coverage civil registration is not, and only the world's richest countries are able to maintain ongoing civil registration systems that cover a majority of the population.  Civil registration in the rest of the world varies from partial coverage to essentially nothing at all \citep{Mathers2005}.  A four-article series titled ``Who Counts?'' in the {\em Lancet} in 2007 reviews the current state of civil registration \citep{AbouZahr2007,Boerma,Hill2007,Horton2007,Mahapatra2007,Setel2007}.  This was followed eight years later with another four-article series presenting a similar but slightly more hopeful picture \citep{abouz2015civil,abouzahr2015towards,mikkelsen2015global,phillips2015well}. The authors lament that there has been a half a century of neglect in civil registration in low- to middle-income countries, and critically, that it is not possible to obtain useful vital statistics from those countries \citep{Mahapatra2007,Setel2007,mikkelsen2015global}.  

The {\em Lancet} authors argue that in the long term all countries need complete civil registration to ensure the civil rights of each one of their citizens and to provide useful, timely public health information \citep{AbouZahr2007,abouzahr2015towards}, and they explore a number of interim options that would allow countries to move from where they are today to full civil registration \citep{Hill2007}.  Echoing the {\em Lancet} special series are additional urgent pleas for better health statistics in low- and middle-income countries \citep[for example: ][]{Abouzahr2010,Bchir2006,Mathers2005,Mathers2009,Rudan2000}.  The WHO and its partners and supporters have actively supported improvements in civil registration and vital statistics (CRVS) over the recent past \citep{world2013strengthening,WHO2013techReportCrvs,WHO2014techReportCrvs}.  These workers clearly identify a  need for representative data describing sex-, age-, and cause-specific mortality through time in small enough areas to be meaningful for local governance and health institutions.  These critiques are for the most part discussed in the framework of civil registration as the `primary' source of data.  

Recently, various United Nations agencies, including the office of the Secretary General, have articulated strong, specific support for rapid improvement in the evidence base for the Sustainable Development Goals (SDG) \citep{sdgsWeb} -- the international target framework that follows from the MDGs \citep[e.g.,][]{dataRevReport,UNdemoEvidence2016,UNRes2016}.  The appropriately named \textit{Data Revolution} \citep{dataRevWeb} is the flagship program organized by the UN to address the systematic lack of data to measure progress toward the SDG targets.

We agree that in order to ensure civil rights and provide each unique citizen with a legal identity, full-coverage civil registration {\em is} the long-term goal. Acknowledging that, we propose decoupling the discussion of civil registration from vital statistics.  In particular, we can obtain accurate and representative vital statistics measurements by making inferences from carefully adjusted samples.


The sample-based approach drives the production of population statistics in many other fields including economics, sociology and political science.  Borrowing from these fields public health workers have developed sample-driven approaches to health statistics that partially substitute for vital statistics derived from civil registration.  India has conducted a sample registration system (SRS) for several decades \citep{indiaSRS} that has produced good basic vital statistics, and more recently Jha and colleagues \citeyearpar{Jha2006} have added verbal autopsy \citep{vaSpecialIssue} to this system to create the Indian Million Death Study (MDS).  In a similar vein, USAID's Sample Vital Registration with Verbal Autopsy (SAVVY) is a program that combines sample registration with verbal autopsy and provides general-purpose tools to collect data \citep{SAVVY}.  USAID's Demographic and Health Surveys (DHS) \citep{DHS} and UNICEF's Multiple Indicator Cluster Surveys (MICS) \citep{MICS} are good examples of traditional household surveys that describe a select subset of indicators for national populations at multiple points in time.  There are many more similar sample surveys conducted by smaller organizations and aimed at specific diseases or the evaluation of specific interventions.  

These approaches generally utilize sampling designs developed to provide cross-sectional snapshots of the current state of the population with respect to an indicator.   With the exception of India's SRS and SAVVY, they lack the ongoing, prospective, longitudinal structure of a traditional vital registration system.  They also often lack the spatial resolution to distinguish differences in indicator values across short distances.  Finally, they often miss or undercount rare events because they typically take one measurement and rely on recall to fill in recent history. 

The current state of the art for public health information in low- to middle-income countries is efforts to combine or triangulate data from multiple sources to produce a more complete picture across both time and space.  The usual sources of data include: non-representative, low-coverage, poor quality vital registration data; roughly once-per-decade census data; snapshot or repeated snapshot data from (sometimes nationally representative) household surveys; one-off sample surveys conducted for a variety of specific reasons by a diverse array of organizations; sample registration systems; and finally, a hodgepodge of miscellaneous data sources that may include health and demographic surveillance systems (HDSS), sentinel surveillance systems, administrative records, clinic/hospital records and others.  

Combining data from different sources with multiple sampling schemes presents a myriad of statistical challenges.  We use \textit{data pooling} as a broad term that describes methods that adjust for bias due to differences in representativeness across data from different sources.  The global burden of disease study by the Institute for Health Metrics and Evaluation \citep{naghavi2015global} is a highly visible example of data pooling.  As another example, \cite{gething:etal:11} pool survey data to produce fine geographical scale Plasmodium falciparum malaria endemicity.  {\em Data amalgamation}, also uses data from multiple sources, but is differentiated by active engagement in the data collection process.  Data amalgamation uses proactive (e.g. {\sc Hyak}) or adaptive mechanisms that actively adjust the data collection process to optimize a set of metrics -- minimize bias, minimize variance, minimize cost, etc.  A recent study of malaria prevalence by \cite{kabaghe2017adaptive} is an example of  {\em data amalgamation} in which survey locations  are adaptively chosen to minimize the variance of a target, see \cite{chipeta:etal:16} for statistical details.
In the survey sampling literature, adaptive cluster sampling has a relatively long history \citep{thompson:seber:96}, and has been used extensively in surveys of rare animal and plant species; we are not aware of any applications in the context considered here.  In short, we use data pooling for situations where researchers combine several datasets not necessarily collected to measure the indicator of interest, whereas data amalgamation is an intentional strategy that incorporates multiple heterogeneous data sources into the design process.
 %

\cite{Rowe2009} and \cite{Bryce2010} describe a system of `integrated, continuous surveys' that would produce ongoing, longitudinal monitoring of a variety of outcomes -- a proactive engagement with the data collection process in keeping with our definition of {\em amalgamation}.  Data from such a system could be representative with respect to population, time and space and thereby substitute for and improve on traditional vital statistics data.  The idea is to systematize the nationally representative household surveys already implemented in a country, conduct them on a regular schedule with a permanent team and institute rigorous quality controls.  The innovation is to turn traditional cross sectional surveys into something quasi longitudinal and to ensure a level of consistency and quality.  This concept appears to still be in the {\em idea} stage without any real methodological development or real-world testing.  More in the spirit of data amalgamation, Bryce and colleagues \citeyearpar{Bryce2004} use a variety of data sources to conduct a multi-country evaluation of Integrated Management of Childhood Illness (IMCI) interventions.  This evaluation does develop some {\em ad-hoc} methods for combining and interpreting data from diverse sources.  

Victora  et al.~\citeyearpar{Victora2011} articulate a similar vision for a national platform for evaluating the effectiveness of public health interventions, specifically those targeting the Millennium Development Goals (MDG).  The authors argue that national coverage with district-level granularity is necessary, and like Rowe and Bryce, that continuous monitoring is required to assess changes and thereby intervention impacts.  That article contains significant discussion of general survey methods, sample size considerations and other methodological requirements that would be necessary to evaluate MDG interventions.  Again however, there are no methodological details that would allow someone to design and implement a national, prospective survey system of the type described.  

Several authors who work at HDSS sites have described an idea for carefully distributing HDSS sites throughout a country in way that could lead to a pseudo representative description of health indicators in the country through time \citep{Ye2012}.  Although these authors do not provide details for how this could be done or evidence that it  works, the basic idea is supported by work from Byass and his colleagues \citeyearpar{Byass2011} who examine the national representativeness of health indicators generated in individual Swedish counties in 1925.  Byass and colleagues discover that any of the not-obviously-unusual counties produced indicator values that were broadly representative of the national population -- the counties being roughly equivalent to an HDSS site, and Sweden in 1925 being roughly equivalent to low- and middle-income countries today.  

Prabhat Jha \citeyearpar{jha2012counting} summarizes all of this in his description of five ideas for improving mortality monitoring with cause of death.  His five ideas include SRS systems with verbal autopsy, improving the representativeness of HDSS (similar to Ye and colleagues \citeyearpar{Ye2012}), coordinating, representative retrospective surveys (similar to Rowe and Bryce) and finally using whatever decent-quality civil registration data might be available.  

We find only two fully implemented and demonstrated examples of data amalgamation in the public health sphere.  Alkema and colleagues \citeyearpar{alkema2007probabilistic,alkema2008bayesian} working with the UNAIDS Reference Group on Estimates, Modelling and Projections develop a Bayesian statistical method that simultaneously estimates the parameters of an epidemiological model that represents the time-evolving dynamics of HIV epidemics {\em and} calibrates the results of that model to match population-wide estimates of HIV prevalence.  The epidemiological model is fit to sentinel surveillance data describing HIV prevalence among pregnant women who attend antenatal clinics, and the population-wide measures of prevalence come from DHS surveys.  Interestingly the second example relates to a similar problem.  Lanjouw and Ivaschenko at the World Bank \citeyearpar{lanjouw2010} describe a method to amalgamate population-level data from DHS surveys and HIV prevalence data from a sentinel surveillance system.  The DHS contains representative information on a variety of items but not HIV prevalence, and the sentinel surveillance system describes the HIV prevalence of a select (non-representative) subgroup, again pregnant women who attend antenatal clinics.  Building on ideas in small-area estimation, they develop and demonstrate a method to adjust the sentinel surveillance data and then predict the HIV prevalence of the whole population.  

Although these are two specific applications of data amalgamation, {\em it is this level of conceptual and methodological detail that are necessary in order to amalgamate data from different sources to produce representative, probabilistically meaningful results.}  The population, public health and evaluation literatures are full of urgent requests for better data and more useful methods to amalgamate data from different sources to answer questions about {\em cause and effect} and {\em change} at national and subnational levels, but there is very little in any of those literatures that actually develops the new concepts and methods that are necessary to deliver the required new capabilities.
\cite{chipeta:etal:16} describe an adaptive design whose aim is to estimate disease prevalence.

\subsection{A New Statistical Platform}

Taking account of the situation described in the literature and firmly in the spirit of `data amalgamation', we aim to develop a system that provides high quality, continuously generated, representative vital statistics and other population and health indicators using a system that is cheap and logistically tractable.  We are confident that such a system can provide highly useful health information at all important geographical (and other) scales: nation, province, district, and perphaps even subdistrict.  

As we argue above, we strongly believe that a {\em sample-based} approach is both appropriate and sufficient to produce meaningful, useful public health information, and we do not believe it is fiscally responsible to attempt to cover the entire population with a public health information system.  That argument must be made on the basis of guaranteeing human rights {\em alone}.

\subsection{Design Criteria}

What we want is a {\em cheap, sustainable}, continuously operated monitoring system that combines the benefits of both sample surveys (representativity, sparse sampling, logistically tractable) and surveillance systems (detailed, linked, longitudinal, prospective with potentially intense monitoring -- e.g. of pregnancy outcomes and neonatal deaths) to provide {\em useful} indicators for large populations over prolonged periods of time, so that we can monitor change and relate changes to possible determinants, including interventions.  More specifically, `useful' in this context means an informative balance of accuracy (bias) and precision (variance) -- i.e. minimal but probably not zero bias accompanied by moderate variance.  {\em We want indicators that are close to the truth most of the time}, and we want an ability to study causality properly.  Critically, we want the whole system to be cheaper and more sustainable than existing systems, and perhaps offer additional advantages as well.

\subsection{{\sc Hyak}}

We propose an integrated data collection and statistical analysis framework for improved population and public health monitoring in areas without comprehensive civil registration and/or vital statistics systems.  We call this platform {\sc Hyak} -- a word meaning `fast' in the Chinook Jargon of the Northwestern United States.  

{\sc Hyak} is conceived as having three fundamental components:

\begin{itemize}
\item {\bf Data Amalgamation}: a sampling and surveillance component that organizes two data collection systems to work together to provide the desired functionality: (1) data from HDSS with frequent, intense, linked, prospective follow-up and (2) data from sample surveys conducted in large areas around the HDSS sites using informed sampling so as to capture as many events as possible.
\item {\bf Verbal Autopsy} \citep{vaSpecialIssue} to estimate the distribution of deaths by cause at the population level, and 
\item {\bf SES}: measurement of socioeconomic status (SES) at household, and perhaps other levels, in order to characterize poverty and wealth.
\end{itemize}

Hyak uses relatively small, intensive, longitudinal HDSS sites to understand what types of individuals (or households) are likely to be the most informative if they were to be included in a sample.  With this knowledge the areas around the HDSS sites are sampled with preference given to the more informative individuals (households), thus increasing the efficiency of sampling and ensuring that sufficient data are collected to describe rare populations and/or rare events.  This fully utilizes the information generated on an on-going basis by the HDSS {\em and} produces indicator values that are representative of a potentially very large area around the HDSS site(s).  Further, the information collected from the sample around the HDSS site can be used to calibrate the more detailed data from the HDSS, effectively allowing the detail in the HDSS data to be extrapolated to the larger population.  For an example of how this has been done in the context of antenatal clinic HIV prevalence surveillance and DHS surveys, see Alkema et al. \citeyearpar{alkema2008bayesian}.  Another way to do this is to build a hierarchical Bayesian model of the indicator of interest, say mortality, with the HDSS being the first (informative) level and the surrounding areas being at the second level.  Thus the surrounding area can borrow information from the HDSS but is not required to match or mirror the HDSS.

%
%
%
%
%

In the remainder of this work we focus on the informed sampling component of {\sc Hyak}.  Informed sampling seeks to capture as many events as possible.  This is critical for the measurement of mortality, and especially for the measurement of cause-specific mortality fractions (CSMF) at the population level.  In order to adequately characterize the epidemiology of a population, it is necessary to measure the CSMF with some precision, and to do this a large number of death events with verbal autopsy are required, especially for rare causes.  Informed sampling aims to make the measurement of mortality rates and CSMFs as efficient as possible.

Below we present a detailed example of the informed sampling idea and a pilot study based on information from the Agincourt HDSS site\footnote{From \cite{kahn2012profile}: The Agincourt health and socio-demographic surveillance system (HDSS), located in rural northeast South Africa close to the Mozambique border, was established in 1992 to support district health systems development led by the post-apartheid ministry of health. At baseline in 1992, 57,600 people were recorded in 8,900 households in 20 villages; by 2006, the population had increased to 70,000 people in 11,700 households. This increase is partly due to Mozambican in-migrants overlooked in the baseline survey and to a new settlement established as part of the post-apartheid governmentÕs Reconstruction and Development Program. In 2007, the study area was extended to include the catchment area of a new privately supported community health centre established to provide HIV treatment before public sector roll-out of HAART. By mid-2011, the population under surveillance comprised 90,000 people residing in 16,000 households in 27 villages. Households are self-defined as Ôpeople who eat from the same pot of foodÕ. Given sustained high levels of temporary labour migration in southern Africa, we included temporary migrants residing for less than 6 months per year who retain close ties with their rural homes in the HDSS. There have been 17 census and vital event update rounds conducted strictly annually since 2000. Participation is virtually complete, with only two households refusing to participate in 2011.} in South Africa \citep{kahn2007research,kahn2012profile}.  The Agincourt HDSS is situated in the rural northeast of South Africa and covers an area of 420km\textsuperscript{2} comprising a sub-district of 27 villages.  The site monitors roughly 90,000 people in 16,000 households.  The villages and households are dispersed widely across this area, and there is a functional road network linking them all.  The epidemiology of the site is typical for South Africa with generally low mortality except for the effect of HIV at very young and middle ages, and in terms of wealth/poverty, the population is typical of a middle-income country \citep[e.g.][]{kabudula2016assessing,clark2015cardiometabolic,houle2014unfolding,clark2013young,gomez2013prevalence,houle2013household}.  The Agincourt HDSS is the canonical HDSS, not extreme along any dimension, and generally representative of what an HDSS site is. 

We generate virtual populations based on information from the Agincourt site, and then we simulate applications of traditional two-stage cluster and {\sc Hyak} sampling designs.  We estimate sex-age-specific mortality rates for children ages $0-4$ years (last birthday) and compare and discuss the results. In the Conclusions Section we describe how verbal autopsy methods can be integrated into the {\sc Hyak} system and the `demographic feasibility' of {\sc Hyak}.  

We are thinking about existing data collection methods and these objectives in a unified framework, and we are starting by experimenting with sampling and analytical frameworks that work together to provide the basis for a {\em measurement system} that is representative, accurate and efficient in terms of information gained per dollar spent (not the same as {\em cheap} in an absolute sense because estimation of a binary outcome like death is still bound by the fundamental constraints of the binomial model; i.e. relatively large numbers of deaths are needed for useful measurements).  

A measurement system like this would be among the cheapest and most informative ways to monitor the mortality of children affected by interventions that cover large areas and exist for prolonged periods of time.  With this in mind, the pilot project we present below focuses on childhood ages $0-4$.


%
%
%

\newpage

\section{Pilot Study of {\sc Hyak} Informed-Sampling via Simulation}

\subsection{Methodological Approach}

In this section we describe our approach to sampling and analysis. To be concrete, we suppose that the outcome of interest is {\em alive} or {\em dead} for children age $0-4$. There are two novel aspects to our approach:%
\begin{itemize}%
\item {\bf Informed Sampling:} Using existing information from a HDSS site we construct a mortality model based on village-level characteristics. On the basis of this model we subsequently predict the number of outcomes of interest in each village of the study region. We then set sample sizes in each village in proportion to these predictions.%
\item {\bf Analysis:} We model the sampled deaths as a function of known demographic factors and village-level characteristics, and then we employ spatial smoothing to tune the model to each village and exploit similarities of risk in neighboring villages.%
\end{itemize}

\subsubsection{Notation}

Given our interest in the binary status {\em alive} or {\em dead}, our modeling framework is logistic regression with random effects. Specifically, let $i=1,\dots,I$ represent villages within the study region and $j=1,\dots,4$ index strata which we take as the four levels of sex (F, M) and age  (Young: $[0,1)$ years, old: $[1,5)$ years). Households within areas will be represented by $k=1, \ldots, K_i$, for $i=1,\dots,I$. The quantity of interest is $Y_{ij}$, the unobserved true number of deaths in village $i$ and in sex/age stratum $j$. We assume that the populations $N_{ij}$ are known in all villages. Also assumed known are village-specific covariates $\bX_{i}$ (for example, the average SES in village $i$, a measure of water quality, or proximity to health care facilities).

The probability of dying in village $i$ and stratum $j$ is denoted by $p_{ij}$, which is the hypothetical proportion of children dying in a hypothetical infinite population in area $i$ and strata $j$. We stress that we are carrying out a small-area estimation problem so the {\em target of interest is $Y_{ij}$} and the probability is just an intermediary which allows us to set up a model. If the full data were observed, we would take the probability to be the observed frequency $\tilde{p}_{ij}=Y_{ij}/N_{ij}$.
The survey design problem corresponds to choosing $n_{ij}$, the number of children in stratum $j$ that we sample in village $i$. Of these, $y_{ij}$ are recorded as dying.

In the next section, we describe models that will be used to analyze the data; once we have estimated probabilities from a generic model, $\widehat{p}_{ij}$, we use the estimator:
\begin{equation}\label{eq:predY}
\widehat{Y}_{ij} = y_{ij} + (N_{ij} - n_{ij}) \times \widehat{p}_{ij},
\end{equation}
where $y_{ij}$ is the observed number of deaths and $(N_{ij}-n_{ij})$ is the number of unsampled individuals in village $i$ and stratum $j$.

\subsubsection{Models}\label{sec:models}

In this section, we describe models that may be fit to the sampled data. 
\begin{enumerate}%
\item[I] {\bf Na\"ive Model:} This baseline model simply estimates $\widehat{p} = y/n$, i.e.,~a single probability is applied to the unsampled individuals in each village. The predicted number of deaths in each village is then (\ref{eq:predY}) with $\widehat{p}_{ij} =\widehat{p}$.%
\item[II] {\bf Strata Model:} This model estimates $\widehat{p}_j = y_j/n_j$, so that estimates of four stratum-specific probabilities are calculated. The predicted number of deaths in each village is then (\ref{eq:predY}) with $\widehat{p}_{ij} =\widehat{p}_j$.%
\item[III] {\bf Covariate Model:} This approach fits a model to data from all villages  where sampling was carried out and estimates stratum effects along with the association between risk and  village-level covariates $\bx_i$. We assume a logistic form,
\begin{equation}\label{eq:LRCM}
  \mbox{logit }p_{ij} =   \bx_{i} \bbeta+  \gamma_j ,
\end{equation}
where $j=1,\dots,4$. Hence, we have a model with a separate baseline for each stratum and with the covariates having a common effect across stratum and village, so there no interaction between covariates and stratum, and covariates and area. We use the maximum likelihood estimates $\widehat{\gamma}_j$ and $\widehat{\bbeta}$ to obtain fitted probabilities: $$\widehat{p}_{ij}= \mbox{expit}(\bx_{i}\widehat{\bbeta}  +  \widehat{\gamma}_j )= \frac{\exp( \bx_{i}\widehat{\bbeta}  +  \widehat{\gamma}_j )}{1+\exp( \bx_{i}\widehat{\bbeta } +  \widehat{\gamma}_j )},$$ which may be used in (\ref{eq:predY}).%
\item[IV] {\bf Spatial Covariate Model:} This approach requires sufficient villages to have sampled data so that spatial random effects can be estimated.  Specifically, we assume a Bayesian implementation of the model:
\begin{equation}\label{eq:LRCMS}
  \mbox{logit }p_{ijk} = \bx_{i}  \bbeta +  \gamma_j + \epsilon_i + S_i + h_k,
\end{equation}
where $j=1,\dots,4$. We have three random effects in this model. The {\it unstructured} village- and household-level error terms $\epsilon_i \sim_{iid} \N(0,\sigma_{\epsilon}^2)$ and $h_k \sim_{iid} \N(0, \sigma_h^2)$, respectively, are independent and allow for excess-binomial variability. The household-level random effects also allow for dependence within households. The $S_i$ error terms are village-level spatial random effects that allow the smoothing of rates across space.  There are many different forms that these random effects could take. A model-based {\it geostatistical approach} \citep{diggle:etal:98} would assume the collection $[S_1,\dots,S_n]$ arise from a multivariate normal distribution, with covariances a function of the distance between villages.  We go a different route and use an intrinsic conditional auto-regressive (ICAR) model \citep{besagbayesian} in which: $$S_i | S_j, j \in \mbox{ne}(i) \sim \N(\overline{S}_i,\sigma_s^2/n_i),$$ where $\mbox{ne}(i)$ is the set of neighbors of village $i$ and $n_i$ is the number of such neighbors. This model assumes that the prior distribution for the spatial effect in area $i$, given its neighbors, is centered on the mean of the neighbors, with a variance that depends on the number of neighbors (with more neighbors reducing the prior variance). We describe our `shared boundary' neighborhood scheme in the next section.  We use the posterior means $\widehat{\bbeta}, \widehat{\gamma}_j, \widehat{\epsilon}_i$ and $\widehat{S}_i$ to obtain fitted probabilities: 
$$\widehat{p}_{ij} =  \frac{\exp( \bx_{i}\widehat{\bbeta}  +  \widehat{\gamma}_j + \widehat{\epsilon}_i + \widehat{S}_i )}{1+\exp(  \bx_{i}\widehat{\bbeta} +  \widehat{\gamma}_j + \widehat{\epsilon}_i + \widehat{S}_i)},$$ 
which may be used in (\ref{eq:predY}); we do not include the household random effects as these are not relevant to predicting an area-level summary, but rather account for within-household clustering. Until relatively recently, fitting this model was computationally challenging within the context of a simulation study (which requires repeated fitting). However, \cite{rue2009approximate} have described a clever combination of Laplace approximations and numerical integration that can be used to carry out Bayesian inference for this this model -- the integrated nested Laplace approximation (INLA). The {\tt INLA} {\tt R} package implements the INLA method. A Bayesian implementation requires specification of priors for all of the unknown parameters, which for model (\ref{eq:LRCMS}) consist of $\bbeta$, $\bgamma$,
$\sigma_{\epsilon}^2$, $\sigma_{s}^2$ and $\sigma_h^{2}$. We choose flat priors for $\bbeta$, $\bgamma$, and Gamma$(a,b)$ priors for $\sigma_{\epsilon}^{-2}$, $\sigma_s^{-2}$ and $\sigma_h^{-2}$.
\end{enumerate}

\begin{figure}[htbp]
\centering
\includegraphics[width=3.7in]{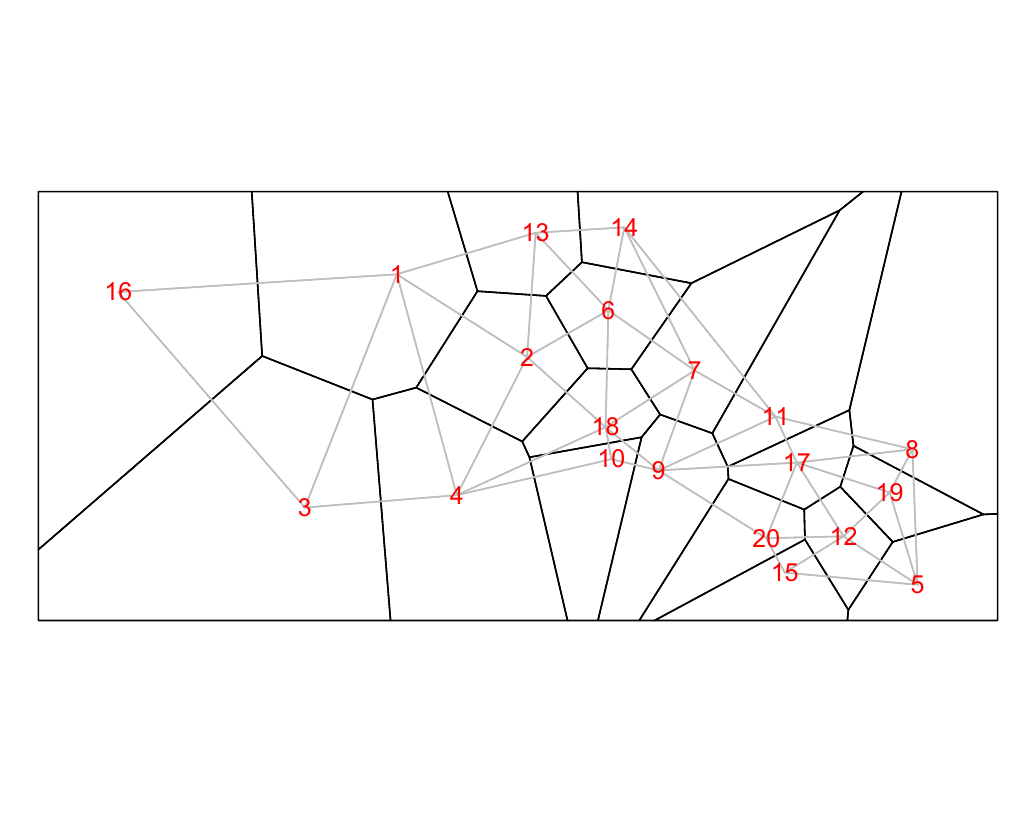}
\caption{The 20 villages of the Agincourt region with Voronoi tesselations defining neighborhood structure. Grey lines indicate neighboring villages.}\label{fig:Villages}
\end{figure}

\subsubsection{The Simulation Study Region}\label{sec:simStudy}

We describe the study region that we create for the simulation study, in order to provide a context within which the different sampling strategies can be described. The study region is based on the Agincourt HDSS site in South Africa \citep{kahn2007research,kahn2012profile}. We assume $N$ individuals reside in one of 20 villages and that there are between 1,400 and 14,000 children in each village, $N_i \sim \mbox{Unif}(1400, 14000)$. In addition for each village, we assume half the children are boys and half are girls, with 20\% in the age range $0-1$ years and 80\% in the age range $1-5$ years. Within each village, we assume that households contain between 1 and 5 children and follow the distribution
\begin{itemize}
\item $P$(household with 1 child) $= 75/470 = 0.16$
\item $P$(household with 2 children) $= 100/470 = 0.21$
\item $P$(household with 3 children) $= 125/470 = 0.27$
\item $P$(household with 4 children) $= 100/470 = 0.21$
\item $P$(household with 5 children) $= 70/470 = 0.15.$
\end{itemize}

We sample a single population of $N$ children and then take $S=100$ repeated draws from this population under the four sampling schemes described below. Beginning with the denominators $N_{ij}$, we sample the observed deaths $y_{ij}$ using a binomial with probabilities given by (\ref{eq:LRCM}).

We sample a second population of $N$ children and treat this population as a historical cohort. It is from this population that we treat three of these villages as HDSS sites for which we have extensive and complete information. 

We form a Voronoi tessellation of the village boundaries based on the 20 coordinate pairs that describe the centroids of the villages. This operation forms a set of tiles, each associated with a centroid and is the set of points nearest to that point. This is a standard operation in spatial statistics \citep[e.g.][]{denison2001bayesian}. We can then define neighbors (for the spatial model) as those villages whose tiles share an edge. Figure \ref{fig:Villages} shows the study region along with village centroids and associated village polygons (as defined by the Voronoi tesselations), along with edges showing the neighborhood structure.

\subsubsection{Sampling Strategies}\label{sec:sampling}

In this section we describe the sampling strategies that we compare. In each strategy, we consider four different sample sizes, $n$, for the total number of children sampled: 1,300, 2,600, 3,900 and 5,200. 
\begin{itemize}%

\item {\bf Two-stage Cluster Sampling:} Randomly select 5 villages and randomly sample $(n/5)/3$ households within each of the villages (since each household contains, on average, 3 children). Additional households will be sampled as needed until at least $n/5$ children are obtained from each village. This is an example of a two-stage cluster sampling plan, a common design. 
\item {\bf Stratified Sampling:} Randomly sample $n/20$ children's outcomes from each of the 20 villages. This strategy lies between the cluster sampling and informed sampling designs.
\item {\bf {\sc Hyak} -- HDSS with Informative Sampling:} The number of children sampled from each village is proportional to the predicted number of deaths based on the HDSS data. In particular, we select all children from the three HDSS villages in the historical cohort and we fit model (\ref{eq:LRCM}). On the basis of the estimated $\bbeta,\bgamma$, we obtain predicted counts of deaths for all villages, using the village-level covariates $\bx_i$, $i=1,\dots,I$. Let $\bbeta^\star,\bgamma^\star$ be the estimated parameters based on the historic HDSS data only and $p_{ij}^{\star}$ be the associated village and stratum-specific probabilities. We estimate $p_{i}^{\star}$ via $$p_{i}^{\star}=\sum_{j=1}^J \frac{N_{ij}}{N_i} p_{ij}^{\star}.$$ Then, the predicted number of deaths are $\widetilde{Y}_{i}=N_{i} \times p^\star_{i}$. We then select  sample sizes as the (rounded versions of) $n_{i} \propto \widetilde{Y}_{i}$ so that villages with more predicted deaths are sampled more heavily. Specifically, we take $n_{i} = n \times \widetilde{Y}_{i}/ \widetilde{Y}_{+}$ where $\widetilde{Y}_{+}$ is the total predicted number of deaths. The observed number of deaths from $n_{i}$ is $y_{i}$. 
\item {\bf Optimum Allocation:} As in the {\sc Hyak} sampling design, we obtain the village-level estimates of the probability of death, $p_{i}^{\star}$, based on the historic HDSS data only. We then select sample sizes as the (rounded versions of) 
$$n_i = n \times \frac{N_i \sqrt{
\widehat{p}_i(1-\widehat{p}_i)
}}{\sum_{i'}N_{i'} \sqrt{
\widehat{p}_{i'}(1-\widehat{p}_{i'}).
}}$$
Details are provided in the Appendix.  

\end{itemize}

\subsubsection{Measures of Predictive Accuracy}

Given $N$ total children, broken into the four stratum, we can set risks $p_{ij}$ (details of which appear in Section \ref{sec:simulation})  for each village/stratum and then simulate counts $Y_{ij}$.  We take this set of $\{ Y_{ij}: i=1,\dots,20;j=1,\dots,4\}$ as fixed, and then subsample from these counts, under each of the four designs and repeat $s=1,\dots,S$ times. 

The estimated number of deaths in survey villages in simulation $s$ is $$\widehat{Y}^{(s)}_{ij} = y^{(s)}_{ij} + (N_{ij}-n_{ij}) \times \widehat{p}^{(s)}_{ij}$$ where the $\widehat{p}^{(s)}_{ij}$ are obtained from one of the models
we described in Section \ref{sec:models}.

To estimate the frequentist properties of the simulation procedure, we summarize the results by examining various summary measures. An obvious measure of accuracy is the mean squared error (MSE) associated with the predicted number of deaths. The MSE of an estimator of the number of deaths in area $i$ and strata $j$, $\widehat{Y}_{ij}$ averaged over villages and strata is
$$\mbox{MSE}( \widehat{Y}_{ij} ) = E \left[ \left(y_{ij}-\widehat{Y}_{ij}  \right)^2 \right],$$
where $y_{ij}$ is the true number of deaths (which recall, is fixed), and the expectation is over all possible samples that can be taken (for whichever design we are considering). This MSE is estimated based on $S$ simulations:
\begin{eqnarray}
\mbox{MSE}( \widehat{\bY})&=& \frac{1}{S}\sum_{s=1}^S \sum_{i=1}^{20} \sum_{j=1}^4  (\widehat{Y}^{(s)}_{ij} - Y_{ij})^2 \nonumber \\
&=& \sum_{i=1}^{20} \sum_{j=1}^4  (
\overline{\widehat{Y}}_{ij}-Y_{ij})^2 + \frac{1}{S}\sum_{s=1}^S \sum_{i=1}^{20} \sum_{j=1}^4  (\widehat{Y}^{(s)}_{ij}-
\overline{\widehat{Y}}_{ij})^2\nonumber \\
&=&  \sum_{i=1}^{20} \sum_{j=1}^4\mbox{Bias}(\widehat{Y}_{ij})^2 + \sum_{i=1}^{20} \sum_{j=1}^4 \mbox{Var}(\widehat{Y}_{ij}).
\label{eqn:mse}
\end{eqnarray}
where $Y_{ij}$ is the true number of deaths in village $i$ and stratum $j$ and $$\overline{\widehat{Y}}_{ij}=\frac{1}{S} \sum_{s=1}^S \widehat{Y}^{(s)}_{ij}$$ is the average of the predicted counts over simulations in village $i$ and stratum $j$. The decomposition in terms of {\bf bias} and {\bf variance} is useful since it makes apparent the trade-off involved in modeling. 

\begin{figure}[htbp]
\centering
\includegraphics[width=4in]{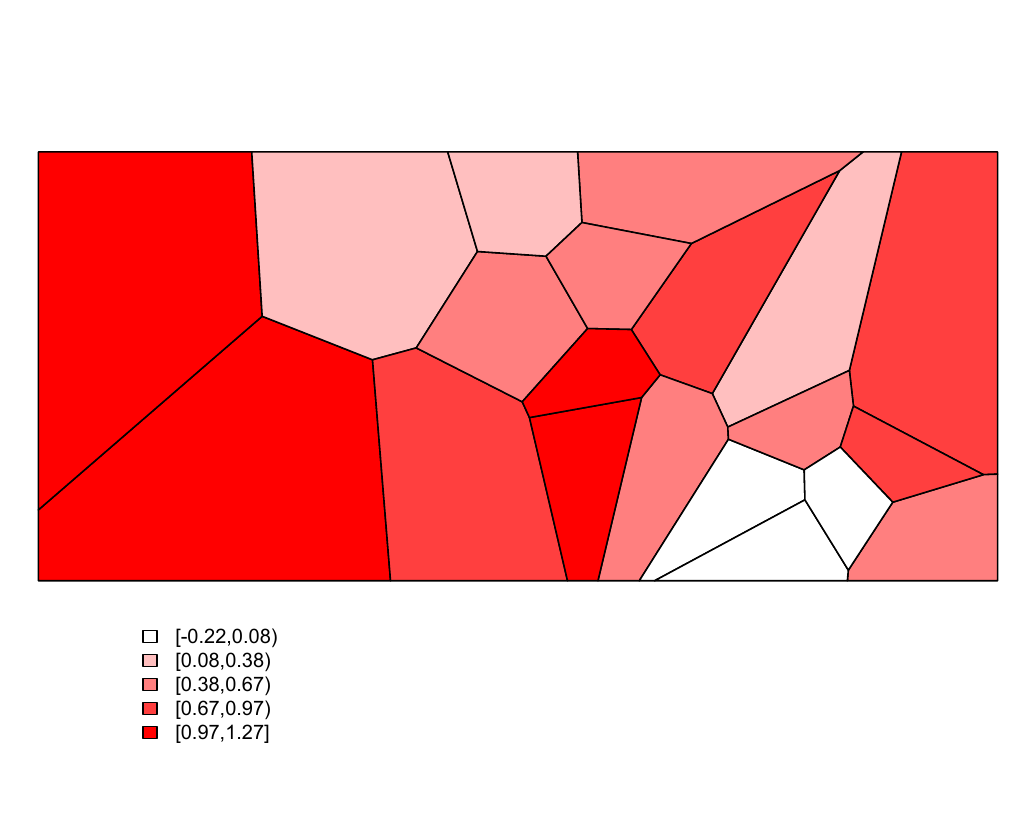}
\caption{The simulated spatial random effects for the Agincourt region.}\label{fig:SpatialEffects}
\end{figure}

\subsection{Simulation}\label{sec:simulation}

We assume there are two village-level covariates so that the length of the $\bbeta$ vector is 2. Both of the village-level covariates $x_{i1}$ and $x_{i2}$ are generated independently from uniform distributions on 0 to 1, $i=1,\dots,20$.  Based loosely on the real values from the Agincourt HDSS in South Africa, the parameter values we use in the simulation are:%
\begin{itemize}%
\item The risk of death in young girls is $\mbox{expit}(\gamma_1) = 0.050$.
%
\item The risk of death in young boys is $\mbox{expit}(\gamma_2) = 0.117$.%
\item The risk of death in older girls is $\mbox{expit}(\gamma_3) = 0.032$.%
\item The risk of death in older boys is $\mbox{expit}(\gamma_4) = 0.077$.%
\item The first village-level covariate has $\exp(\beta_1) = \exp(-2.2) = 0.111$ so that a unit increase in $x_1$ leads to the odds of death dropping by one ninth. 
\item The second village-level covariate has $\exp(\beta_2) = \exp(1.4) = 4.05 $ so that a unit increase in $x_2$ leads to the odds of death quadrupling.
\item We set $\sigma_{\epsilon}^2 = 0.22$ to determine the level of unstructured variability at the village level. This leads to a 95\% range for the residual unstructured village-level odds being $\exp( \pm 1.96 \times \sqrt{0.22}) = [0.40,2.51]$.%
\item We set $\sigma_{s}^2 = 0.48$ to determine the level of structured variability at the village level. This operation requires some care because the ICAR model does not define a proper probability distribution. The ICAR variance is not interpretable as a marginal variance (and so is not comparable to the other random effects variances, $\sigma_\epsilon^2$ and $\sigma_h^2$) and so instead Figure \ref{fig:SpatialEffects} shows a simulated set of $S_i$, $i=1,\dots,20$ values, with darker values indicating higher risk. The spatial dependence is apparent, with this realization producing high risk to the West of the region and low risk in the East.%

\item We set $\sigma_h^2 = 0.08$ to determine the level of unstructured variability at the household level. This leads to a 95\% range for the residual unstructured household-level odds being $\exp( \pm 1.96 \times \sqrt{0.08}) = [0.57, 1.74]$.
\end{itemize}

For the strata and covariates model, the covariate relationship is estimated from the villages that produced data, and then model (\ref{eq:LRCM}) is used to obtain fitted probabilities that are applied to the unsampled villages, using the population and covariate information that is assumed known for each village.

\begin{figure}[htbp]
\centering
\subfigure[]{
\includegraphics[height=2.75in,width=2.75in]{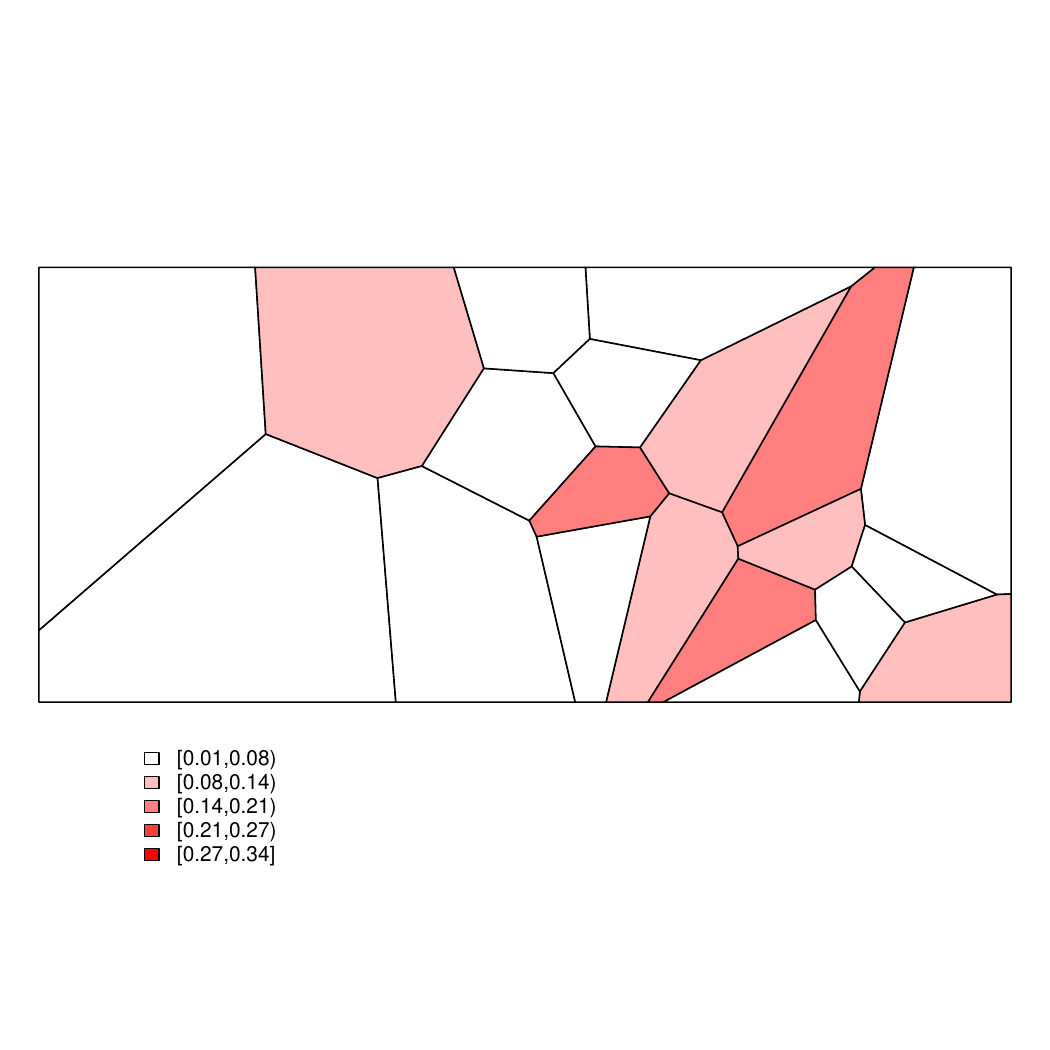}}
\subfigure[]{
\includegraphics[height=2.75in,width=2.75in]{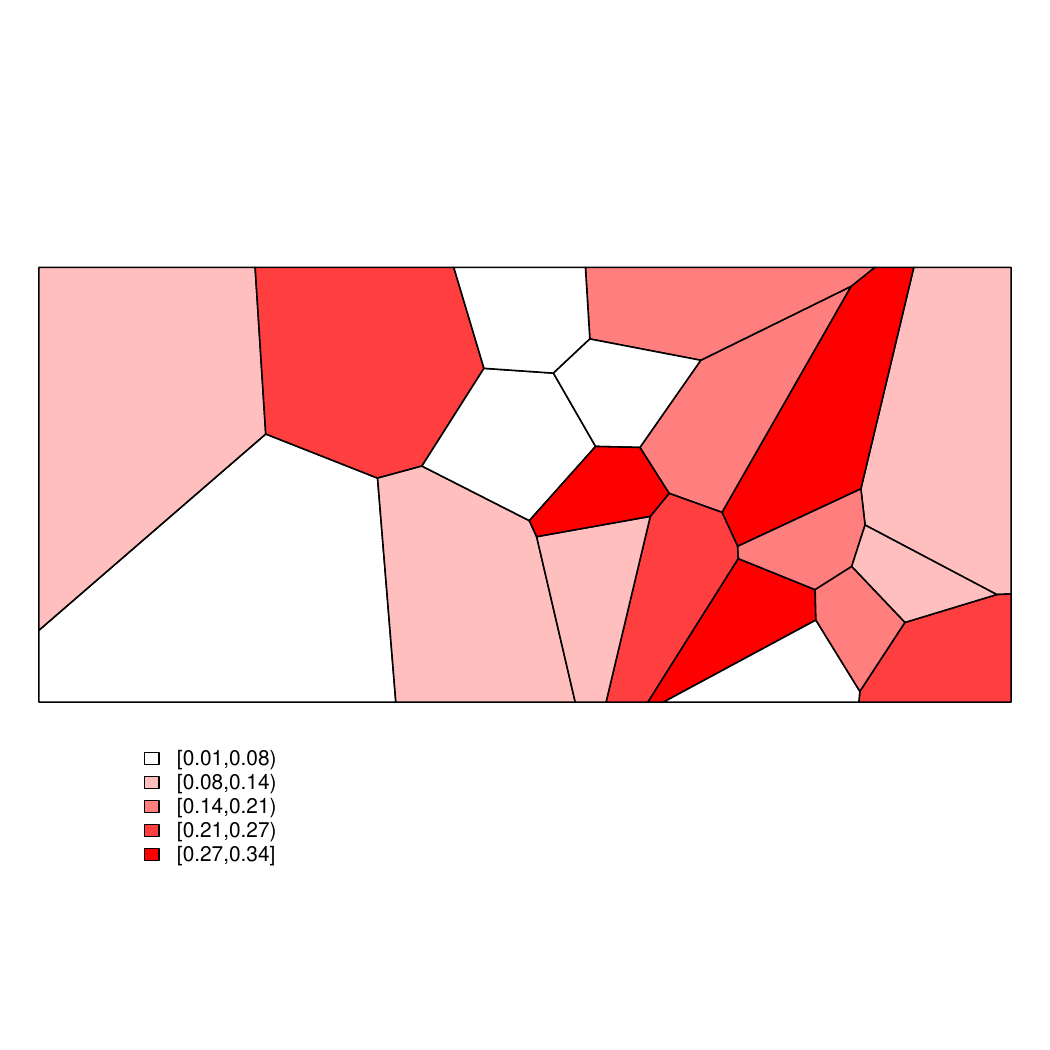}}
\subfigure[]{
\includegraphics[height=2.75in,width=2.75in]{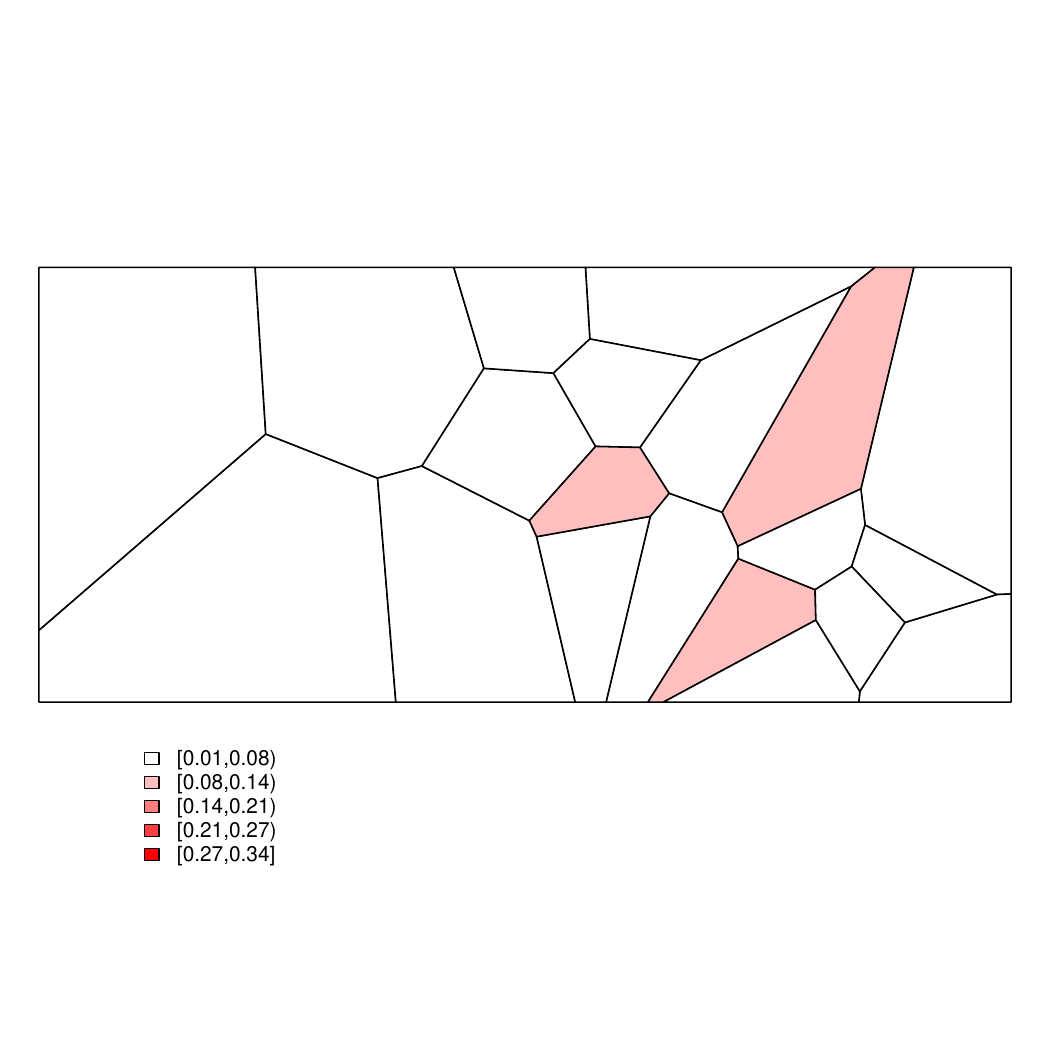}}
\subfigure[]{
\includegraphics[height=2.75in,width=2.75in]{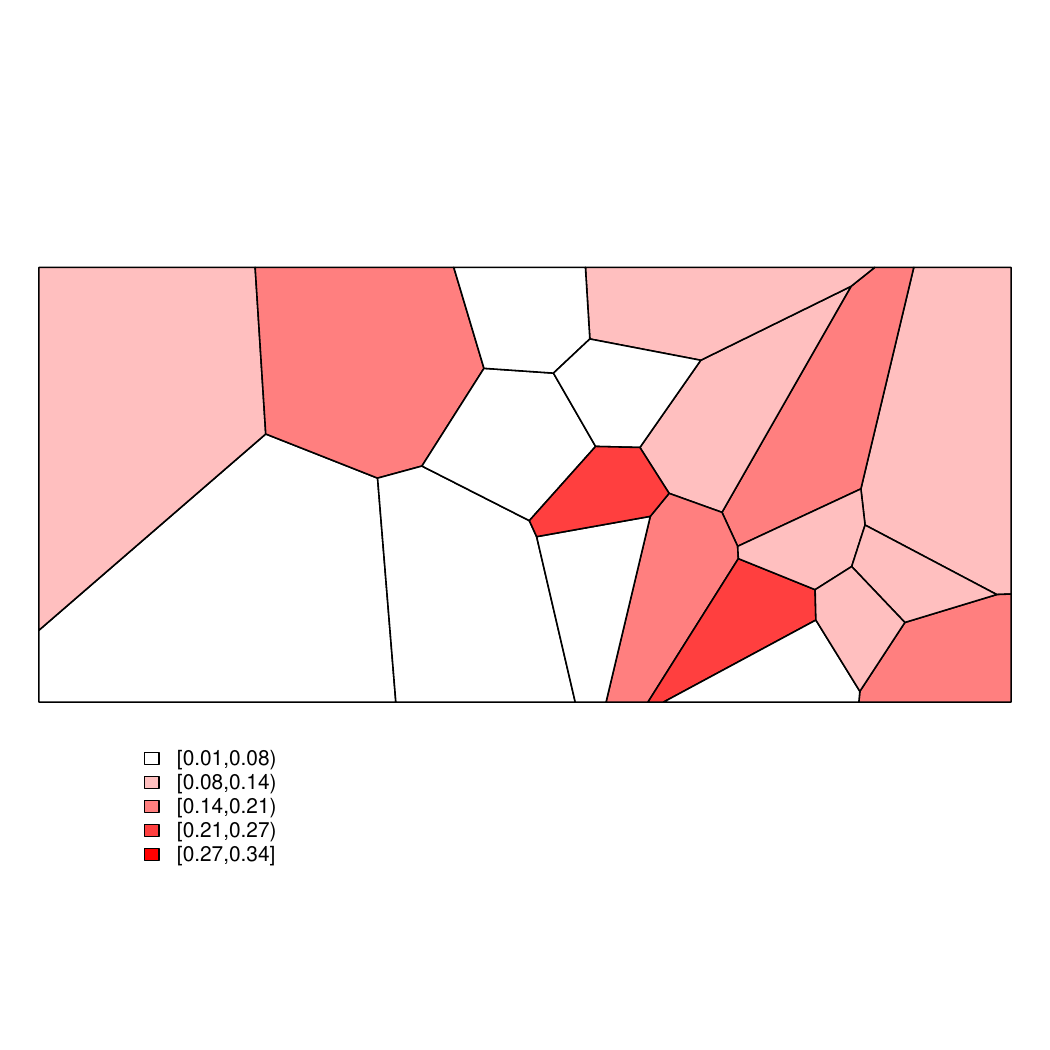}}
\caption{The predicted probabilities of dying for the Agincourt region: (a) young girls, (b) young boys, (c) older girls, (d) older boys.}\label{fig:stratumprobs}
\end{figure}

Combining all of the elements of the model, we generate deaths $Y_{ij}$ for village $i$ and stratum $j$ by randomly drawing from a Binomial distribution with probabilities given by (\ref{eq:LRCMS}). This yields the predicted probabilities for all 20 villages and for each of the four stratum displayed in Figure \ref{fig:stratumprobs}. The historic cohort is generated in the same fashion. Details of the village-level characteristics for both cohorts are provided in Appendix A.2. 

The HDSS villages are selected by taking the villages with both large $x_1$ and large $x_2$, small $x_1$ and small $x_2$, followed by a randomly sampled third village. 

 A $Ga(5,1)$ prior is used for the spatial and non-spatial random effects in the spatial models (Model IV).

\subsection{Results}\label{sec:results}

Table \ref{onepop} summarizes the results of the simulation study for $n=5,200$. Results for the smaller sample sizes are shown in Tables \ref{onepop:n3900}-\ref{onepop:n1300} in Appendix A.3 The number of average sampled deaths and bias, variance and MSE from (\ref{eqn:mse}) are displayed for each combination of sampling strategy and analytical model.  


Overall, the {\sc Hyak} sampling strategy captures more deaths and is generally more accurate.  Across sampling schemes and sample sizes, {\sc Hyak} generally has the smallest MSEs. Further examination of the components of the MSE reveals that: (i) {\sc Hyak} yields smaller bias, and (ii) pays for this by sacrificing some variance.  The overall comparison between the sampling strategies clearly favors {\sc Hyak}.  This partly reflects the careful choice of HDSS villages so that they contain substantial variation in terms of village-level covariates.

\begin{center}
\begin{minipage}{6.1in}
\begin{threeparttable}[H]
\centering
  \caption{Deaths, Bias, Variance, MSE for cluster sampling, stratified sampling, {\sc Hyak} and optimum sampling for $n=5,200$. Results from $S=100$ simulations. There were 11,299 deaths in the simulated population from which samples were taken. `Cluster' is shorthand for {\em Two-stage Cluster Sample};  `{\sc Hyak}' for {\em HDSS with Informative Sampling}; `Strata/Covariates' for {\em Logistic Regression Covariate Model} and `Strata/Covariates/Space' for {\em Logistic Regression Random Effects Covariate Model}. It is not possible to fit the spatial model (IV) to the two-stage cluster sampling scheme since there are data from 5 villages only.}
    \begin{tabular}{rlcccc}
    \toprule
    \multicolumn{1}{l}{Design} & \multicolumn{1}{l}{Model} & \multicolumn{1}{c}{Deaths} & \multicolumn{1}{c}{Bias} & \multicolumn{1}{c}{Variance ($\times 10^3$)} & \multicolumn{1}{c}{MSE ($\times 10^3$)} \bigstrut[b]\\
    \midrule 
    \multicolumn{1}{c}{\multirow{4}[0]{*}{Cluster}} & I. Na\"{i}ve & 459 & 1,067 & 174 & 1,312  \bigstrut[t]\\
    \multicolumn{1}{c}{} & II. Strata & 459 & 874 & 188 & 951 \\
    \multicolumn{1}{c}{} & III. Strata/Covariates& 459 & 651 & 386 & 810 \\
    \multicolumn{1}{c}{} & IV. Strata/Covariates/Space & 459 & --- & --- & --- \\
       \midrule 
    \multicolumn{1}{c}{\multirow{4}[0]{*}{Stratified}} & I. Na\"{i}ve & 460 & 1,058 & 5 & 1,124 \bigstrut[t]\\ 
    \multicolumn{1}{c}{} & II. Strata & 460 & 866 & 15 & 765 \\
    \multicolumn{1}{c}{} & III. Strata/Covariates & 460 & 651 & 16 & 439 \\
    \multicolumn{1}{c}{} & IV. Strata/Covariates/Space & 460 & 183 & 80 & 113  \\
    \midrule
    \multicolumn{1}{c}{\multirow{4}[0]{*}{Hyak}} & I. Na\"{i}ve & 538 & 1,162 & 7 & 1,357 \bigstrut[t]\\
    \multicolumn{1}{c}{} & II. Strata & 538 & 969 & 18 & 956 \\
    \multicolumn{1}{c}{} & III. Strata/Covariates & 538 & 635 & 16 & 419 \\
    \multicolumn{1}{c}{} & IV. Strata/Covariates/Space & 538 & 182 & 66 & \textcolor{red}{100} \\
    \midrule
    \multicolumn{1}{c}{\multirow{4}[0]{*}{Optimum}} & I. Na\"{i}ve & 477 & 1,072 & 5 & 1,154 \\
    \multicolumn{1}{c}{} & II. Strata & 477 & 880 & 18 & 792 \\
    \multicolumn{1}{c}{} & III. Strata/Covariates & 477 & 632 & 17 & 416 \\
    \multicolumn{1}{c}{} & IV. Strata/Covariates/Space & 477 & 167 & 74 & 102 \bigstrut[b]\\
    \bottomrule
    \end{tabular}%
\label{onepop}
\end{threeparttable}
\end{minipage}
\end{center}

Comparing the analytical models also produces an encouraging result.  Within each sampling strategy, the logistic regression random effects covariate model (model IV) performs best overall (smaller MSEs). Within {\sc Hyak}, this outperforms the others. Similar patterns are observed across all sample sizes. This suggests that accounting for unmeasured factors and taking advantage of the spatial structure of mortality risk is significantly worthwhile. 

\vspace{0.5cm}

\begin{figure}[htbp]
\centering
\includegraphics[width=5in]{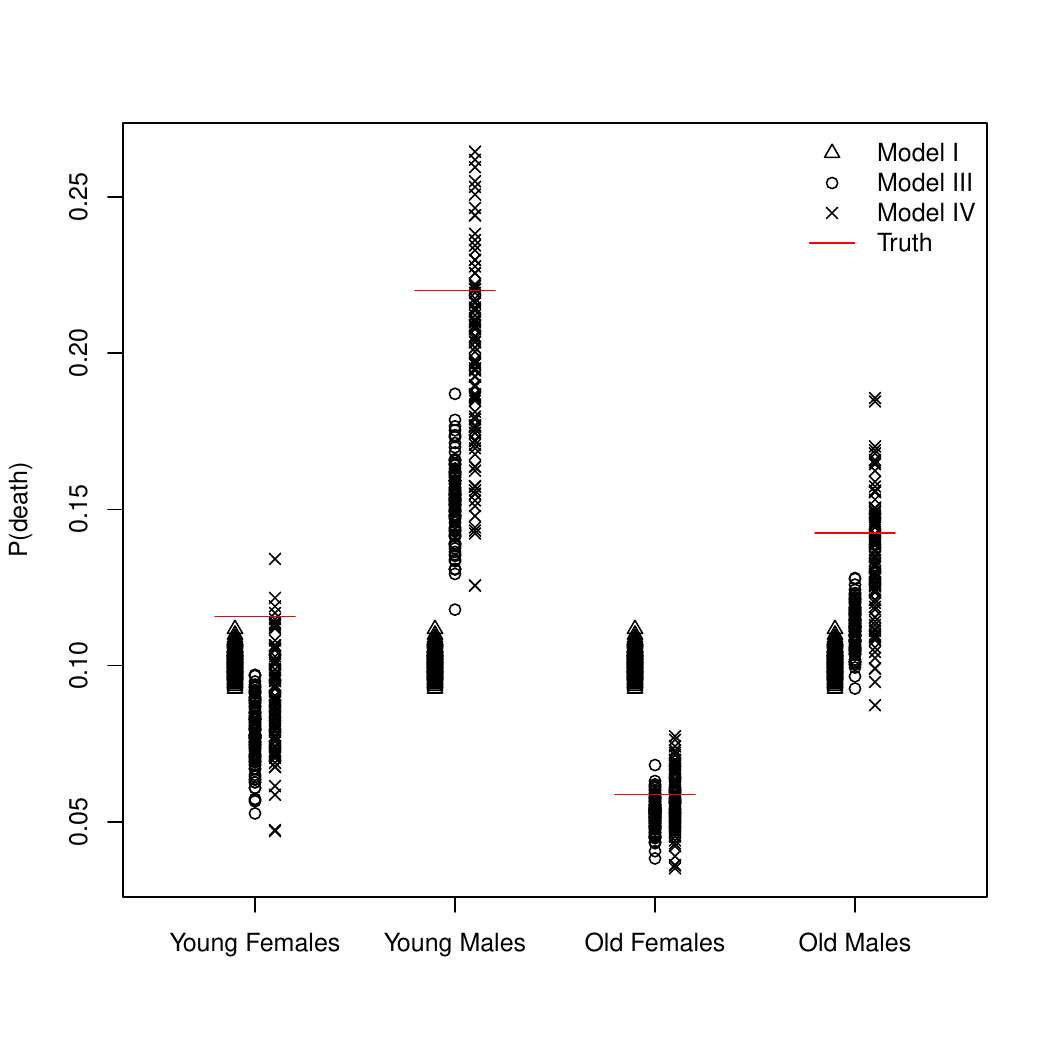}
\caption{The distributions of the estimated probability of dying from models I, III and IV under the {\sc Hyak} sampling strategy for $n=5,200$.}\label{fig:tradeOff_n5200}
\end{figure}

The trade-off between bias and variance is clearly revealed by a closer look at the distributions of the estimated probability of dying produced by each model.  Figure \ref{fig:tradeOff_n5200} displays these distributions for models I, III \& IV -- {\em Na\"{i}ve, Covariates} and {\em Covariates \& Space} under the {\sc Hyak} sampling strategy for $n=5,200$, while figures \ref{fig:tradeOff_n3900}-\ref{fig:tradeOff_n1300} in Appendix A.3 display these same distributions for $n=3,900, n=2,600$ and $n=1,300$, respectively.  The {\em Na\"{i}ve} model estimates are very condensed, always miss the truth and have clear bias; estimates from the {\em Covariates} model also have very little spread, almost always miss the truth and have some bias; and finally, estimates from the {\em Covariates \& Space} model have large spread, however the distributions nearly always include the truth, and have much less bias.  Clearly the {\em Covariates \& Space} model displays the balance we are seeking: small bias and manageable spread, and importantly, distributions that include the truth.  This combination of sampling strategy and analytical approach provides our key objective: an indicator that is close to (and around) the truth most of the time. 

\begin{figure}[htbp]
\centering
\includegraphics[width=5in]{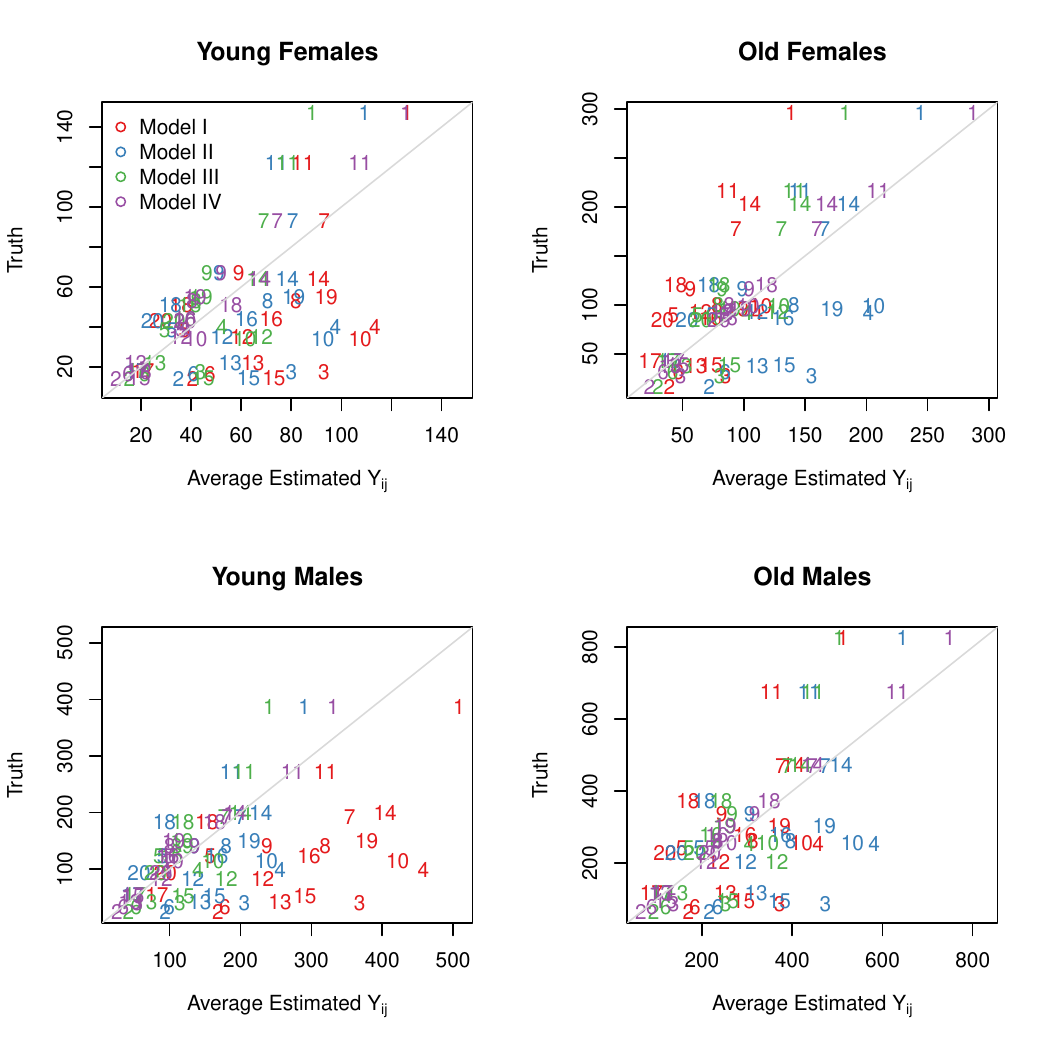}
\caption{The average village- and strata-specific estimates for the (unobserved) population counts of death plotted against the true values across each of the four models under the {\sc Hyak} sampling scheme for $n=5,200$. Plotting symbols indicate village numbers, and colors indicate model number with key in upper-left plot.  The spatial model IV (purple) symbols are in general closest to the $y=x$ line of equality.}
\label{fig:Ests_vs_Truth_n5200}
\end{figure}

Figure \ref{fig:Ests_vs_Truth_n5200} displays the average village- and strata-specific estimates for the (unobserved) population counts of death plotted against the true values across each of the four models under the {\sc Hyak} sampling scheme for $n=5,200$, while figures \ref{fig:Ests_vs_Truth_n3900}-\ref{fig:Ests_vs_Truth_n1300} in Appendix A.3 display the same for the smaller sample sizes. (See figures \ref{fig:Cluster_Ests_vs_Truth_n5200}-\ref{fig:Optim_Ests_vs_Truth_n1300} in Appendix A.3 for the remaining sampling schemes for all sample sizes.) In general, the average estimates from the spatial model tend to follow the $y=x$ line quite closely, indicating we are estimating the true number of deaths in each village quite well. Estimates tend to be closer under the {\sc Hyak} sampling strategy and for larger sample sizes, thus confirming (visually) our previous results.


\section{Discussion}

\subsection{Key Conclusions}

The key conclusion of this pilot study is that the statistical sampling and analysis ideas supporting the {\sc Hyak}  monitoring system are sound: a combination of highly informative data such as are produced by a HDSS site can be used to judiciously inform sampling of a large surrounding area to yield estimated counts of deaths that are far more useful than those produced by a traditional cluster sample design.  Further, {\sc Hyak} combined with an analytical model that includes unstructured random effects and spatial smoothing produces the most accurate and well-behaved estimates.  The improvements are dramatic and clearly justify additional work on these ideas.

Another crucial idea underlying {\sc Hyak} is the notion that very detailed information generated by an HDSS site can be extrapolated to the much larger surrounding population by calibrating that information with carefully chosen and much less detailed data from the surrounding population.  This idea has already been demonstrated convincingly by Alkema et al. [\citeyear{alkema2008bayesian}] and is currently being applied by UNAIDS to produce global estimates of HIV prevalence.  This relies on the assumption that the population monitored by the HDSS is similar enough to the population surrounding the HDSS that the relationships between covariates and the outcomes of interest are the same or very similar.  The degree to which this is true will vary among specific settings.  In particular, when HDSS sites also serve as research and intervention testing sites, it is possible that there will be \textit{Hawthorne Effect} issues -- i.e. the intensively studied HDSS population will be different from the surrounding population that has not participated in studies and trials.  This may affect the key covariate-outcome relationships that drive {\sc Hyak}.  This is something that must be studied, initially with a real-world pilot study of  {\sc Hyak}, and then in an ongoing way by occasionally verifying these relationships through an oversample of the surrounding population, or through small add-on studies conducted whenever a census is done in the surrounding areas to update the sampling frame.  Although this is a concern, it is unlikely to make {\sc Hyak} infeasible or invalidate  {\sc Hyak} results.  An explicit goal of a pilot study will be to characterize the uncertainty created by possible Hawthorne Effect issues and build them into  {\sc Hyak} estimates.

A key advantage of {\sc Hyak} sampling strategy is that it {\em captures significantly more deaths}.  Verbal autopsy methods \citep{vaSpecialIssue} can be applied to all or a fraction of these deaths to assign causes (immediate, contributing, etc.).  This cause of death information can then be used to construct distributions of deaths by cause -- CSMFs -- which illuminate the epidemiological regime affecting the population, and if this is monitored through time, how the epidemiology of the population is changing.  Critically, this provides a means of measuring the impact of interventions on specific causes of death and the distribution of deaths over time.  The increased number of deaths captured with informed sampling increases the accuracy and precision of measurements of CSMFs.

A final benefit of the {\sc Hyak} system is that it provides two types of infrastructure: the HDSS and the sample survey.  In addition to providing information with which to sample, the HDSS provides a platform on which a wide variety of longitudinal studies can be undertaken -- linked observational studies; randomized, controlled trials, all kinds of combinations of these, etc.  Moreover, the permanent HDSS infrastructure also provides a training platform that can support a wide variety of health and behavioral science training, mentoring and apprenticing/interning and experience for young scientists or health professionals.  Having the sample survey infrastructure provides a means of quickly validating/calibrating studies conducted by the HDSS and provides another learning dimension for the educational and training activities that the system can support.

A potential limitation of any mortality monitoring system is `demographic feasibility', that is the ability to capture enough deaths in a given population to measure levels and/or changes in mortality, potentially by cause, through time.  Death is a binomial process defined by a probability of dying, and as such, is governed by the characteristics of the binomial model. That model specifies in simple terms the number of deaths necessary to estimate the probability of dying within a given margin of error with a given level of confidence.  No amount of sophistication will release us from that basic set of facts.  The {\sc Hyak} system addresses this challenge by providing a means through which to choose the best possible sample given what we know about the population, and this in turn maximizes our ability to capture deaths.  The fundamentals of the binomial model require that one must observe relatively large numbers of deaths to measure mortality precisely and especially to measure changes in mortality with both precision and confidence.  So in light of those inescapable realities, the {\sc Hyak} system produces the most information per dollar spent, because it captures more deaths per dollar spent.

Finally and perhaps most importantly, the {\sc Hyak} monitoring system is cheaper to run over a period of years compared to traditional cluster sample-based survey methods.  Combined with the fact that {\sc Hyak} also produces more useful information, this makes {\sc Hyak} highly cost effective -- {\em more bang for less buck.}

Importantly, there are implementation considerations that must be addressed before {\sc Hyak} can be used at provincial or national scale to provide population-representative estimates.  These will need to be resolved through additional theoretical work, simulation, and ultimately through a pilot study that conducts {\sc Hyak} on a large population dispersed over a large physical space.  Among many, these critical questions need to be answered:
\begin{itemize}
\item How big do HDSS sites need to be to provide enough information for effective informative sampling?
\item How many HDSS sites are necessary for effective informed sampling with respect to key demographic and epidemiological indicators?
\item How should HDSS sites be dispersed geographically?
\item How well does  {\sc Hyak} work to provide disaggregated (fine-grained) estimates of key indicators by sex, age, wealth/poverty, space, time, etc?
\item How much does the sampling frame affect  {\sc Hyak} results, and what cheap, feasible solutions are there to obtaining frequently updated sampling frames?
\item A detailed costing and cost comparison needs to be done comparing the costs of the the HDSS site; the additional census, sampling, and interviewing needed for {\sc Hyak}; and a traditional household multi-stage cluster sample survey (like DHS) conducted in the same area.
\item How the method can be scaled up to a larger geographical area. We envisage that only a subset of villages will be sampled, and then a geostatistical model \citep{wakefield:simpson:godwin:16} can be used for spatial prediction to unobserved villages (a critical question is the number of villages needed to train the spatial model). Another important issue is to deal with the potential problem of preferential sampling \citep{diggle:etal:10} in which sampling locations are selected based on the expected size of the response. In order to inform sampling historical data  (for example, DHS surveys) may be used to model to create a predictive surface, upon which sampling may be based. Investigating this idea will be the subject of a future paper.
\end{itemize}

\newpage
\bibliographystyle{apalike}
\bibliography{hyak.bib}

\newpage
\appendix
\numberwithin{table}{section}
\numberwithin{figure}{section}
\section{Appendix}
\subsection{Optimum allocation sampling strategy details}

Suppose we have stratum, indexed by $i=1,\dots,I$. In our case the strata are areas. Let $N_i$ be the population of area $i$ and $N=\sum_i N_i$ the total population in the study region.

Let $Y_{ik}=0/1$ be the indicator of whether child $k$ in area $i$ died, $k=1,\dots,N_i$, $i=1,\dots,I$. Then we are interested in $T=\sum_i\sum_k Y_{ik}$, the total number of deaths. The fraction of deaths is $\overline{y}=\widehat{p}=T/N$.

Let $q_i=N_i/N$ and $S_i$ be the standard deviation of the response in stratum $i$ where $$S^2_i = \frac{N_i}{N_i-1}
{p}_i(1-{p}_i) \approx {p}_i(1-{p}_i)
,$$ which is estimated by
$$s^2_i = \frac{n_i}{n_i-1}
\widehat{p}_i(1-\widehat{p}_i) \approx \widehat{p}_i(1-\widehat{p}_i)
.$$

If we use the usual estimator of $\widehat{p}_i=\sum_{k=1}^{n_i} y_{ik}/n_i$ then the variance is
$$var( \overline{y}) = \sum_{i=1}^I q_i^2 (1-f_i) \frac{S_i^2}{n_i} =  \sum_{i=1}^I q_i^2 (1-f_i)\frac{N_i}{N_i-1}
\frac{{p}_i(1-{p}_i)}{n_i},$$
where $f_i=n_i/N_i$,
which leads to
$$var(\widehat{T} ) =N^2 \sum_{i=1}^I q_i^2(1-f_i)  \frac{S_i^2}{n_i} = N^2\sum_{i=1}^Iq_i^2 (1-f_i) \frac{p_i(1-p_i)}{n_i-1}.$$
Substituting in $\widehat{p}_i$ gives the estimated variances.

We wish to choose $n_i$, the number of samples to take in area $i$.

Then the optimum allocation, in the sense of minimizing $var(\overline{y})$ (which is the same as minimizing the variance of $T$) is Neyman allocation in which
\begin{equation}\label{eq:alloc}
n_i =n \frac{q_i S_i}{\sum_i q_i S_i}.
\end{equation}
Note: we really should be minimizing MSE as our estimators are biased (since they are random effects models with shrinkage).

In our setting, we have an estimate of $p_i$ and so we can use this in (\ref{eq:alloc}) which becomes
\begin{equation}\label{eq:alloc2}
n_i \approx n \times \frac{q_i \sqrt{
\widehat{p}_i(1-\widehat{p}_i)
}}{\sum_{i'}q_{i'} \sqrt{
\widehat{p}_{i'}(1-\widehat{p}_{i'}).
}}.
\end{equation}
We do not include the age-gender groups $j$ in our sampling strata, but our model produces estimates $\widehat{p}_{ij}$ so we estimate $\widehat{p}_i$ via
$$
\widehat{p}_i = \sum_{j=1}^J \frac{N_{ij}}{N_i} \widehat{p}_{ij},
$$
to use in (\ref{eq:alloc2}).

\subsection{Village-level characteristics for the current and historic cohorts}

Tables \ref{tab:currentcohort} and \ref{tab:historicalcohort} display the village characteristics for both the current-day and historical cohorts. The current-day cohort is the fixed population from which we draw repeated samples, while the historical cohort is used by the {\sc Hyak} and optimum sampling schemes to obtain estimated village-level probabilities of death. In our simulation, we used villages 4, 7 and 8 as the HDSS sites.
 
\begin{table}[h]
\caption{Village characteristics for current-day cohort. This cohort represents our fixed population from which we draw repeated samples.}
\begin{center}
\begin{tabular}{ccccccc}
\hline
Village & Number of Households & Number of Children & \# Deaths & P(Death) & $\bx_1$ & $\bx_2$ \\ 
  \hline
1 & 4221 & 12523 & 1654 & 0.13 & 0.56 & 0.70 \\ 
  2 & 1376 & 4150 & 119 & 0.03 & 0.92 & 0.32 \\ 
  3 & 3050 & 9172 & 169 & 0.02 & 0.89 & 0.55 \\ 
  4 & 3804 & 11331 & 483 & 0.04 & 0.92 & 0.56 \\ 
  5 & 1275 & 3802 & 492 & 0.13 & 0.39 & 0.68 \\ 
  6 & 1515 & 4550 & 156 & 0.03 & 0.58 & 0.17 \\ 
  7 & 3036 & 9011 & 929 & 0.10 & 0.77 & 0.98 \\ 
  8 & 2648 & 7870 & 554 & 0.07 & 0.32 & 0.07 \\ 
  9 & 1957 & 5841 & 658 & 0.11 & 0.55 & 0.83 \\ 
  10 & 3532 & 10630 & 500 & 0.05 & 0.57 & 0.47 \\ 
  11 & 2679 & 7981 & 1286 & 0.16 & 0.10 & 0.60 \\ 
  12 & 2034 & 6043 & 413 & 0.07 & 0.05 & 0.83 \\ 
  13 & 2082 & 6291 & 218 & 0.03 & 0.73 & 0.17 \\ 
  14 & 3320 & 9901 & 939 & 0.09 & 0.76 & 0.96 \\ 
  15 & 2466 & 7361 & 196 & 0.03 & 0.53 & 0.51 \\ 
  16 & 2467 & 7301 & 531 & 0.07 & 0.66 & 0.44 \\ 
  17 & 709 & 2092 & 230 & 0.11 & 0.04 & 0.51 \\ 
  18 & 1192 & 3610 & 725 & 0.20 & 0.02 & 0.76 \\ 
  19 & 3083 & 9300 & 600 & 0.06 & 0.62 & 0.27 \\ 
  20 & 836 & 2482 & 447 & 0.18 & 0.09 & 0.97 \\ 
   \hline
\end{tabular}
\end{center}
\label{tab:currentcohort}
\end{table}%

\begin{table}[h]
\caption{Village characteristics for historical cohort. The HDSS villages are 4, 7 and 8.}
\begin{center}
\begin{tabular}{ccccccc}
\hline
Village & Number of Households & Number of Children & \# Deaths & P(Death) & $\bx_1$ & $\bx_2$ \\ 
  \hline
1 & 1460 & 4331 & 587 & 0.14 & 0.56 & 0.70 \\ 
  2 & 4064 & 12001 & 331 & 0.03 & 0.92 & 0.32 \\ 
  3 & 524 & 1552 & 33 & 0.02 & 0.89 & 0.55 \\ 
  4 & 2927 & 8720 & 377 & 0.04 & 0.92 & 0.56 \\ 
  5 & 4022 & 11891 & 1499 & 0.13 & 0.39 & 0.68 \\ 
  6 & 4157 & 12450 & 393 & 0.03 & 0.58 & 0.17 \\ 
  7 & 2873 & 8532 & 919 & 0.11 & 0.77 & 0.98 \\ 
  8 & 1529 & 4540 & 322 & 0.07 & 0.32 & 0.07 \\ 
  9 & 4108 & 12152 & 1292 & 0.11 & 0.55 & 0.83 \\ 
  10 & 1570 & 4640 & 231 & 0.05 & 0.57 & 0.47 \\ 
  11 & 2789 & 8342 & 1444 & 0.17 & 0.10 & 0.60 \\ 
  12 & 3685 & 10931 & 693 & 0.06 & 0.05 & 0.83 \\ 
  13 & 1786 & 5242 & 165 & 0.03 & 0.73 & 0.17 \\ 
  14 & 674 & 2070 & 187 & 0.09 & 0.76 & 0.96 \\ 
  15 & 473 & 1402 & 31 & 0.02 & 0.53 & 0.51 \\ 
  16 & 3187 & 9550 & 735 & 0.08 & 0.66 & 0.44 \\ 
  17 & 4344 & 13080 & 1329 & 0.10 & 0.04 & 0.51 \\ 
  18 & 3449 & 10302 & 2058 & 0.20 & 0.02 & 0.76 \\ 
  19 & 3080 & 9191 & 666 & 0.07 & 0.62 & 0.27 \\ 
  20 & 468 & 1422 & 286 & 0.20 & 0.09 & 0.97 \\ 
   \hline
\end{tabular}
\end{center}
\label{tab:historicalcohort}
\end{table}

\newpage
\subsection{Additional simulation results}
Tables \ref{onepop:n3900}, \ref{onepop:n2600}, and \ref{onepop:n1300} summarize the results of the simulation study for $n=3,900, n=2,600$ and $n=1,300$, respectively. The number of average sampled deaths and bias, variance and MSE from (\ref{eqn:mse}) are displayed for each combination of sampling strategy and analytical model. 

\begin{center}
\begin{minipage}{6.1in}
\begin{threeparttable}[h]
\centering
  \caption{Deaths, Bias, Variance, MSE for cluster sampling, stratified sampling, {\sc Hyak} and optimum sampling for $n=3,900$. Results from $S=100$ simulations. There were 11,299 deaths in the simulated population from which samples were taken. `Cluster' is shorthand for {\em Two-stage Cluster Sample};  `{\sc Hyak}' for {\em HDSS with Informative Sampling}; `Strata/Covariates' for {\em Logistic Regression Covariate Model} and `Strata/Covariates/Space' for {\em Logistic Regression Random Effects Covariate Model}. It is not possible to fit the spatial model (IV) to the two-stage cluster sampling scheme since there are data from 5 villages only.}
    \begin{tabular}{rlcccc}
    \toprule
    \multicolumn{1}{l}{Design} & \multicolumn{1}{l}{Model} & \multicolumn{1}{c}{Deaths} & \multicolumn{1}{c}{Bias} & \multicolumn{1}{c}{Variance ($\times 10^3$)} & \multicolumn{1}{c}{MSE ($\times 10^3$)} \bigstrut[b]\\
    \midrule 
    \multicolumn{1}{c}{\multirow{4}[0]{*}{Cluster}} & I. Na\"{i}ve & 342 & 1,072 & 192 & 1,342  \bigstrut[t]\\
    \multicolumn{1}{c}{} & II. Strata & 342 & 878 & 207 & 977 \\
    \multicolumn{1}{c}{} & III. Strata/Covariates& 342 & 644 & 775 & 1,190 \\
    \multicolumn{1}{c}{} & IV. Strata/Covariates/Space & 342 & --- & --- & --- \\
       \midrule 
    \multicolumn{1}{c}{\multirow{4}[0]{*}{Stratified}} & I. Na\"{i}ve & 344 & 1,066 & 9 & 1,145 \bigstrut[t]\\ 
    \multicolumn{1}{c}{} & II. Strata & 344 & 871 & 26 & 785 \\
    \multicolumn{1}{c}{} & III. Strata/Covariates & 344 & 660 & 25 & 460 \\
    \multicolumn{1}{c}{} & IV. Strata/Covariates/Space & 344 & 225 & 99 & 150  \\
    \midrule
    \multicolumn{1}{c}{\multirow{4}[0]{*}{Hyak}} & I. Na\"{i}ve & 409 & 1,181 & 8 & 1,402 \bigstrut[t]\\
    \multicolumn{1}{c}{} & II. Strata & 409 & 982 & 25 & 988 \\
    \multicolumn{1}{c}{} & III. Strata/Covariates & 409 & 640 & 22 & 431 \\
    \multicolumn{1}{c}{} & IV. Strata/Covariates/Space & 409 & 188 & 92 & 128 \\
    \midrule
    \multicolumn{1}{c}{\multirow{4}[0]{*}{Optimum}} & I. Na\"{i}ve & 356 & 1,079 & 7 & 1,171 \\
    \multicolumn{1}{c}{} & II. Strata & 356 & 885 & 23 & 806 \\
    \multicolumn{1}{c}{} & III. Strata/Covariates & 356 & 642 & 23 & 436 \\
    \multicolumn{1}{c}{} & IV. Strata/Covariates/Space & 356 & 194 & 85 & \textcolor{red}{123} \bigstrut[b]\\
    \bottomrule
    \end{tabular}%
\label{onepop:n3900}
\end{threeparttable}
\end{minipage}
\end{center}

\begin{center}
\begin{minipage}{6.1in}
\begin{threeparttable}[h]
\centering
  \caption{Deaths, Bias, Variance, MSE for cluster sampling, stratified sampling, {\sc Hyak} and optimum sampling for $n=2,600$. Results from $S=100$ simulations. There were 11,299 deaths in the simulated population from which samples were taken. `Cluster' is shorthand for {\em Two-stage Cluster Sample};  `{\sc Hyak}' for {\em HDSS with Informative Sampling}; `Strata/Covariates' for {\em Logistic Regression Covariate Model} and `Strata/Covariates/Space' for {\em Logistic Regression Random Effects Covariate Model}. It is not possible to fit the spatial model (IV) to the two-stage cluster sampling scheme since there are data from 5 villages only.}
    \begin{tabular}{rlcccc}
    \toprule
    \multicolumn{1}{l}{Design} & \multicolumn{1}{l}{Model} & \multicolumn{1}{c}{Deaths} & \multicolumn{1}{c}{Bias} & \multicolumn{1}{c}{Variance ($\times 10^3$)} & \multicolumn{1}{c}{MSE ($\times 10^3$)} \bigstrut[b]\\
    \midrule 
    \multicolumn{1}{c}{\multirow{4}[0]{*}{Cluster}} & I. Na\"{i}ve & 250 & 1,075 & 170 & 1,326  \bigstrut[t]\\
    \multicolumn{1}{c}{} & II. Strata & 250 & 881 & 190 & 966 \\
    \multicolumn{1}{c}{} & III. Strata/Covariates& 250 & 659 & 382 & 816 \\
    \multicolumn{1}{c}{} & IV. Strata/Covariates/Space & 250 & --- & --- & --- \\
       \midrule 
    \multicolumn{1}{c}{\multirow{4}[0]{*}{Stratified}} & I. Na\"{i}ve & 256 & 1,075 & 11 & 1,166 \bigstrut[t]\\ 
    \multicolumn{1}{c}{} & II. Strata & 256 & 879 & 30 & 802 \\
    \multicolumn{1}{c}{} & III. Strata/Covariates & 256 & 664 & 27 & 468 \\
    \multicolumn{1}{c}{} & IV. Strata/Covariates/Space & 256 & 248 & 123 & 185  \\
    \midrule
    \multicolumn{1}{c}{\multirow{4}[0]{*}{Hyak}} & I. Na\"{i}ve & 302 & 1,193 & 15 & 1,439 \bigstrut[t]\\
    \multicolumn{1}{c}{} & II. Strata & 302 & 992 & 41 & 1,025 \\
    \multicolumn{1}{c}{} & III. Strata/Covariates & 302 & 646 & 30 & 448 \\
    \multicolumn{1}{c}{} & IV. Strata/Covariates/Space & 302 & 209 & 109 & \textcolor{red}{152} \\
    \midrule
    \multicolumn{1}{c}{\multirow{4}[0]{*}{Optimum}} & I. Na\"{i}ve & 264 & 1,090 & 10 & 1,198 \\
    \multicolumn{1}{c}{} & II. Strata & 264 & 893 & 31 & 829 \\
    \multicolumn{1}{c}{} & III. Strata/Covariates & 264 & 646 & 29 & 446 \\
    \multicolumn{1}{c}{} & IV. Strata/Covariates/Space & 264 & 223 & 109 & 159 \bigstrut[b]\\
    \bottomrule
    \end{tabular}%
\label{onepop:n2600}
\end{threeparttable}
\end{minipage}
\end{center}

\begin{center}
\begin{minipage}{6.1in}
\begin{threeparttable}[h]
\centering
  \caption{Deaths, Bias, Variance, MSE for cluster sampling, stratified sampling, {\sc Hyak} and optimum sampling for $n=1,300$. Results from $S=100$ simulations. There were 11,299 deaths in the simulated population from which samples were taken. `Cluster' is shorthand for {\em Two-stage Cluster Sample};  `{\sc Hyak}' for {\em HDSS with Informative Sampling}; `Strata/Covariates' for {\em Logistic Regression Covariate Model} and `Strata/Covariates/Space' for {\em Logistic Regression Random Effects Covariate Model}. It is not possible to fit the spatial model (IV) to the two-stage cluster sampling scheme since there are data from 5 villages only.}
    \begin{tabular}{rlcccc}
    \toprule
    \multicolumn{1}{l}{Design} & \multicolumn{1}{l}{Model} & \multicolumn{1}{c}{Deaths} & \multicolumn{1}{c}{Bias} & \multicolumn{1}{c}{Variance ($\times 10^3$)} & \multicolumn{1}{c}{MSE ($\times 10^3$)} \bigstrut[b]\\
    \midrule 
    \multicolumn{1}{c}{\multirow{4}[0]{*}{Cluster}} & I. Na\"{i}ve & 113 & 1,079 & 193 & 1,358  \bigstrut[t]\\
    \multicolumn{1}{c}{} & II. Strata & 113 & 886 & 241 & 1,025 \\
    \multicolumn{1}{c}{} & III. Strata/Covariates& 113 & 662 & 1,252 & 1,690 \\
    \multicolumn{1}{c}{} & IV. Strata/Covariates/Space & 113 & --- & --- & --- \\
       \midrule 
    \multicolumn{1}{c}{\multirow{4}[0]{*}{Stratified}} & I. Na\"{i}ve & 119 & 1,088 & 23 & 1,205 \bigstrut[t]\\ 
    \multicolumn{1}{c}{} & II. Strata & 119 & 895 & 62 & 863 \\
    \multicolumn{1}{c}{} & III. Strata/Covariates & 119 & 662 & 60 & 499 \\
    \multicolumn{1}{c}{} & IV. Strata/Covariates/Space & 119 & 325 & 196 & 301  \\
    \midrule
    \multicolumn{1}{c}{\multirow{4}[0]{*}{Hyak}} & I. Na\"{i}ve & 138 & 1,193 & 24 & 1,447 \bigstrut[t]\\
    \multicolumn{1}{c}{} & II. Strata & 138 & 1,001 & 70 & 1,071 \\
    \multicolumn{1}{c}{} & III. Strata/Covariates & 138 & 655 & 61 & 491 \\
    \multicolumn{1}{c}{} & IV. Strata/Covariates/Space & 138 & 309 & 175 & \textcolor{red}{271} \\
    \midrule
    \multicolumn{1}{c}{\multirow{4}[0]{*}{Optimum}} & I. Na\"{i}ve & 122 & 1,100 & 27 & 1,238 \\
    \multicolumn{1}{c}{} & II. Strata & 122 & 902 & 78 & 891 \\
    \multicolumn{1}{c}{} & III. Strata/Covariates & 122 & 658 & 68 & 500 \\
    \multicolumn{1}{c}{} & IV. Strata/Covariates/Space & 122 & 306 & 203 & 297 \bigstrut[b]\\
    \bottomrule
    \end{tabular}%
\label{onepop:n1300}
\end{threeparttable}
\end{minipage}
\end{center}

\begin{figure}[htbp]
\centering
\includegraphics[width=5in]{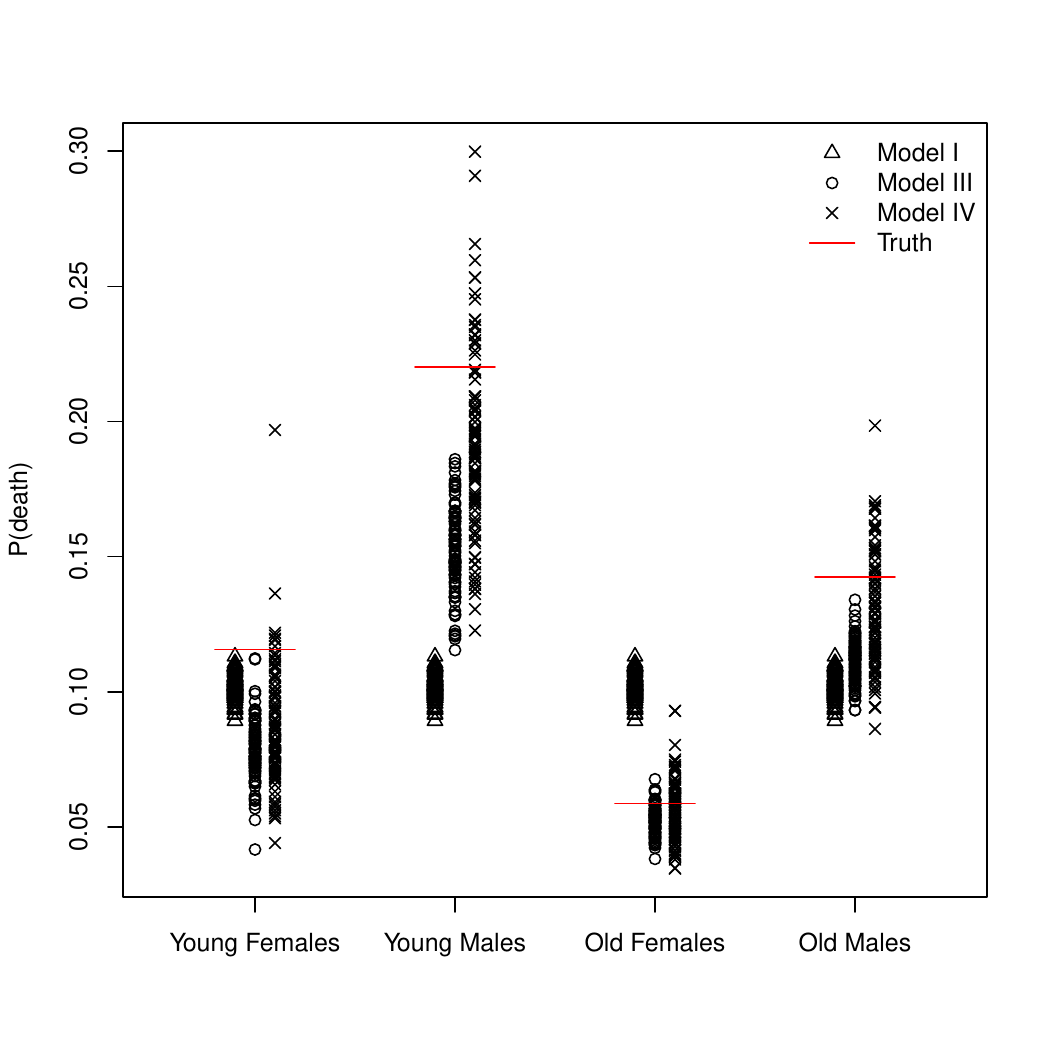}
\caption{The distributions of the estimated probability of dying from models I, III and IV under the {\sc Hyak} sampling strategy for $n=3,900$.}\label{fig:tradeOff_n3900}
\end{figure}

\begin{figure}[htbp]
\centering
\includegraphics[width=5in]{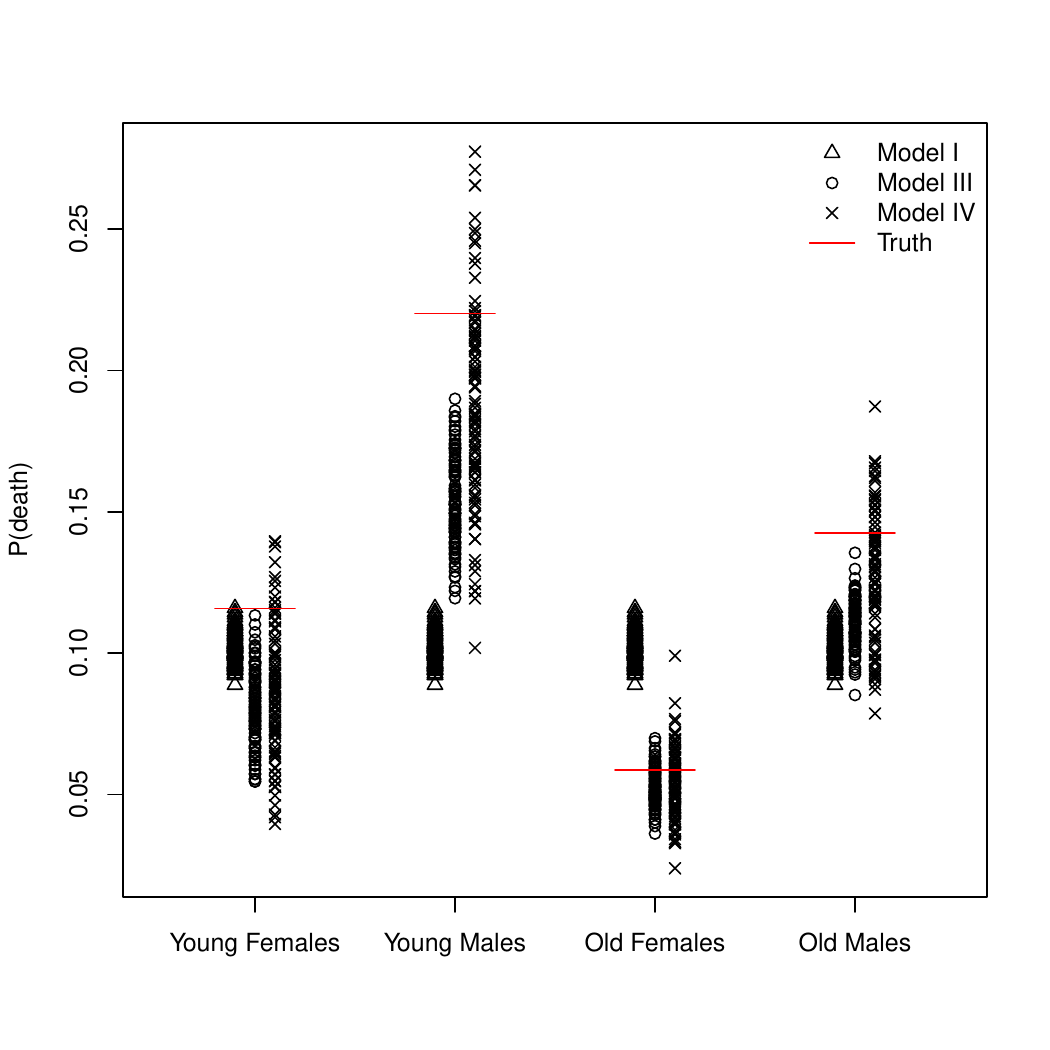}
\caption{The distributions of the estimated probability of dying from models I, III and IV under the {\sc Hyak} sampling strategy for $n=2,600$.}\label{fig:tradeOff_n2600}
\end{figure}

\begin{figure}[htbp]
\centering
\includegraphics[width=5in]{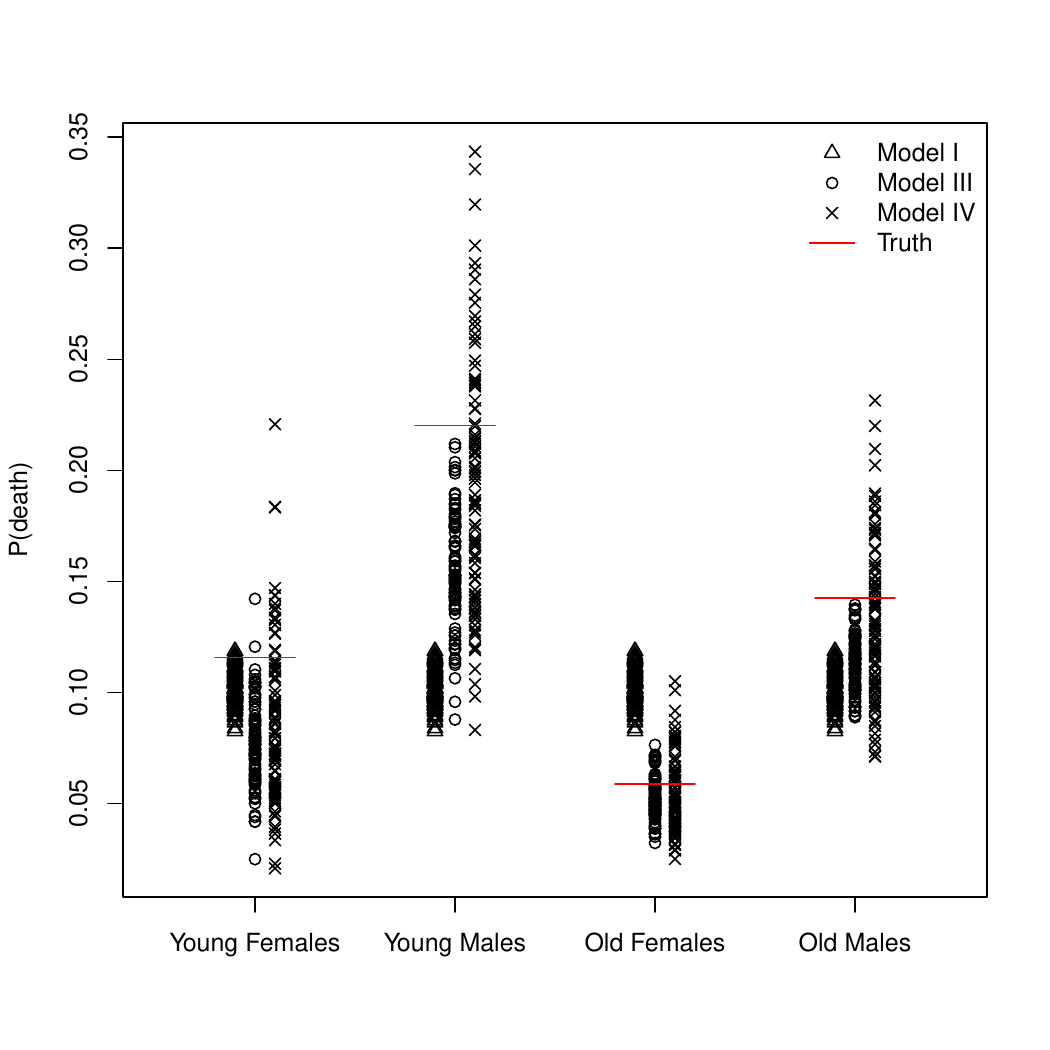}
\caption{The distributions of the estimated probability of dying from models I, III and IV under the {\sc Hyak} sampling strategy for $n=1,300$.}\label{fig:tradeOff_n1300}
\end{figure}

\clearpage
Figures \ref{fig:tradeOff_n3900}-\ref{fig:tradeOff_n1300} display the distributions of the estimated probability of dying produced by each model (models I, III \& IV -- {\em Na\"{i}ve, Covariates} and {\em Covariates \& Space}) under the {\sc Hyak} sampling strategy for $n=3,900, n=2,600$ and $n=1,300$, respectively.

\begin{figure}[htbp]
\centering
\includegraphics[width=5in]{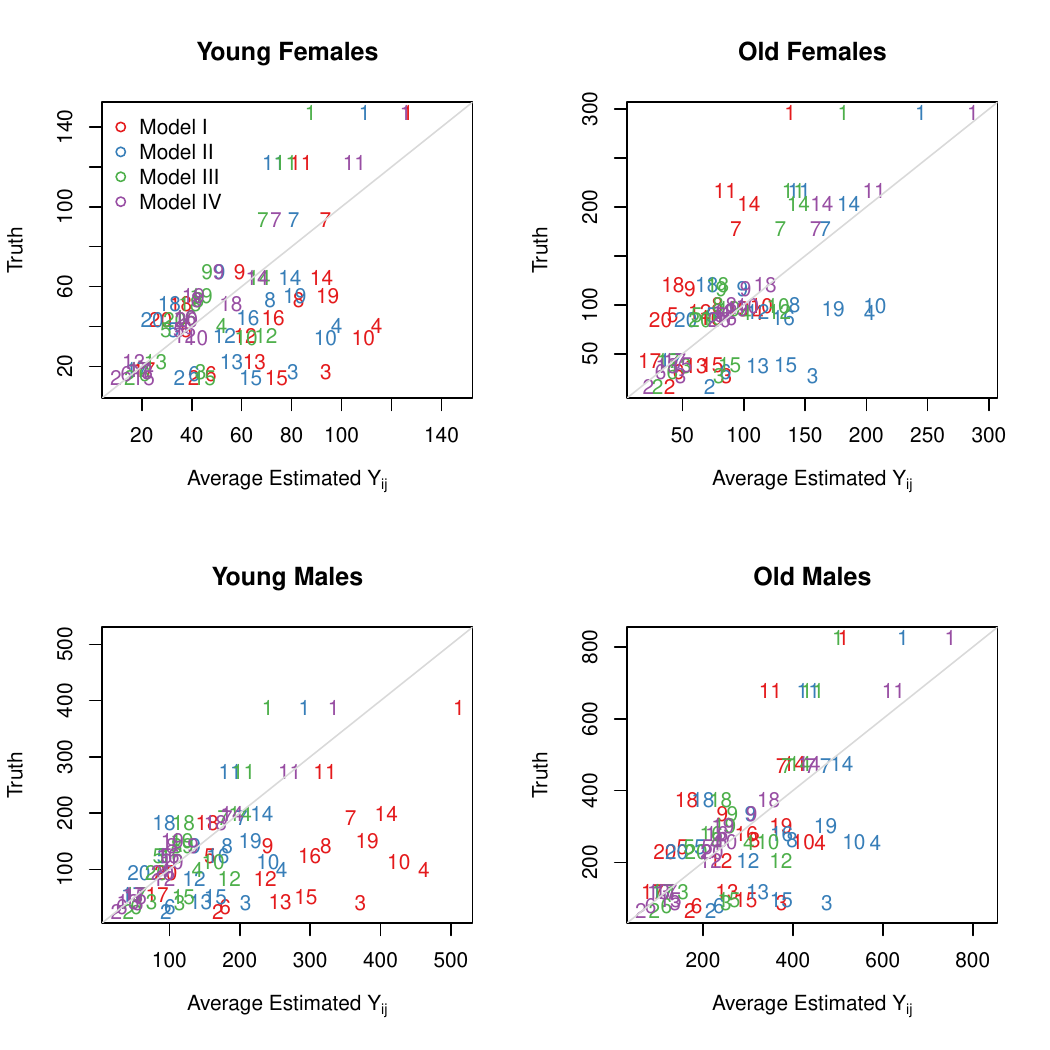}
\caption{The average village- and strata-specific estimates for the (unobserved) population counts of death plotted against the true values across each of the four models under the {\sc Hyak} sampling scheme for $n=3,900$. Plotting symbols indicate village numbers.}
\label{fig:Ests_vs_Truth_n3900}
\end{figure}

\begin{figure}[htbp]
\centering
\includegraphics[width=5in]{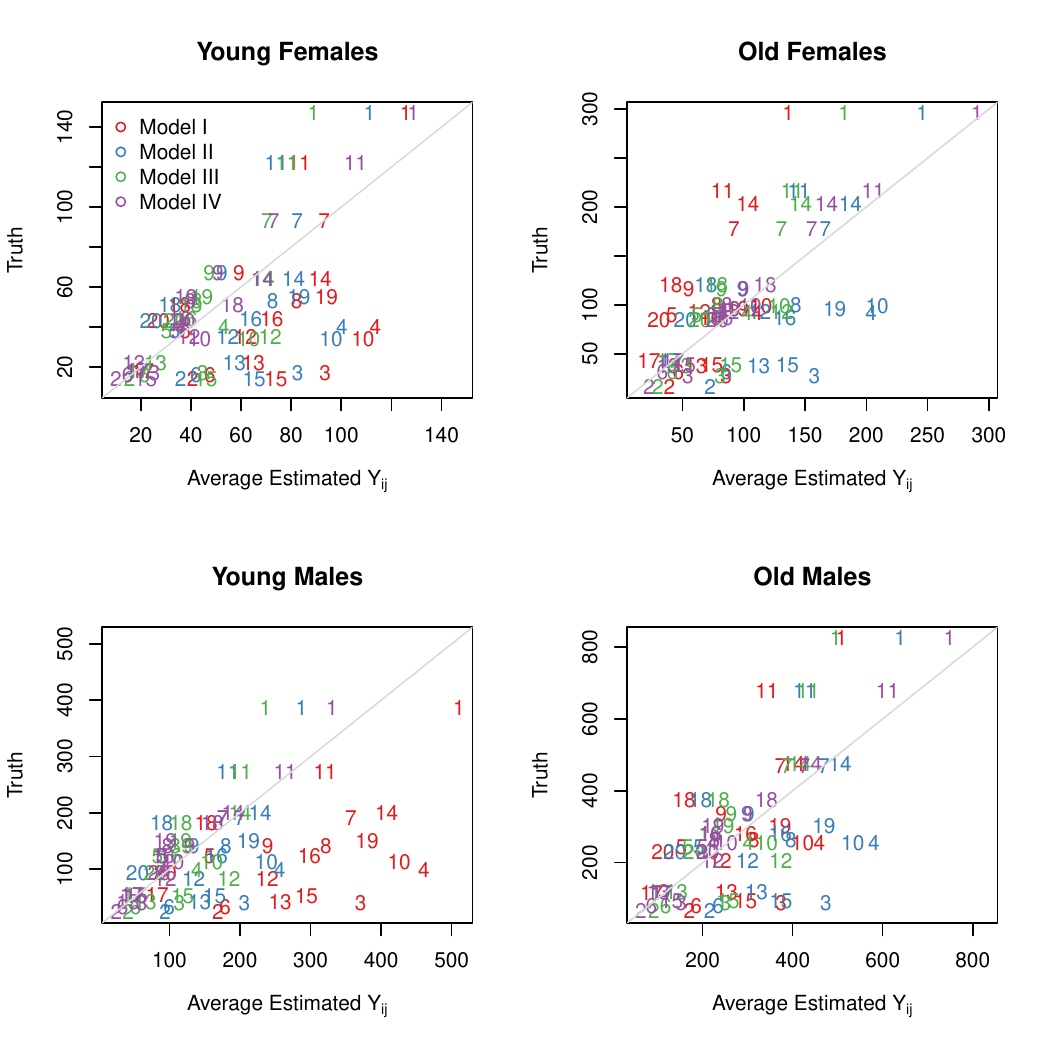}
\caption{The average village- and strata-specific estimates for the (unobserved) population counts of death plotted against the true values across each of the four models under the {\sc Hyak} sampling scheme for $n=2,600$. Plotting symbols indicate village numbers.}
\label{fig:Ests_vs_Truth_n2600}
\end{figure}

\begin{figure}[htbp]
\centering
\includegraphics[width=5in]{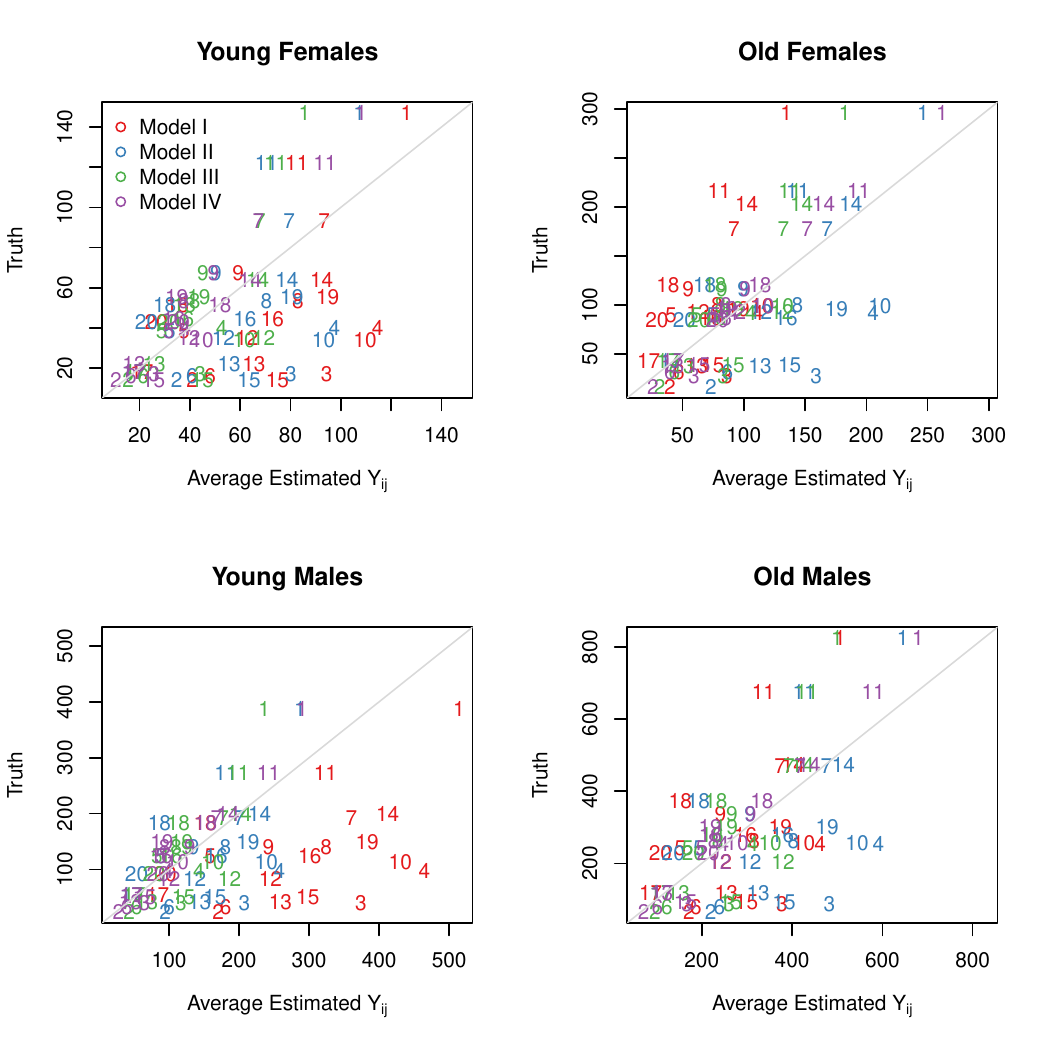}
\caption{The average village- and strata-specific estimates for the (unobserved) population counts of death plotted against the true values across each of the four models under the {\sc Hyak} sampling scheme for $n=1,300$. Plotting symbols indicate village numbers.}
\label{fig:Ests_vs_Truth_n1300}
\end{figure}

\clearpage
Figures \ref{fig:Ests_vs_Truth_n3900}-\ref{fig:Ests_vs_Truth_n1300} display the average village- and strata-specific estimates for the (unobserved) population counts of death plotted against the true values across each of the four models under the {\sc Hyak} sampling scheme for $n=3,900, n=2,600$ and $n=1,300$, respectively.

\begin{figure}[htbp]
\centering
\includegraphics[width=5in]{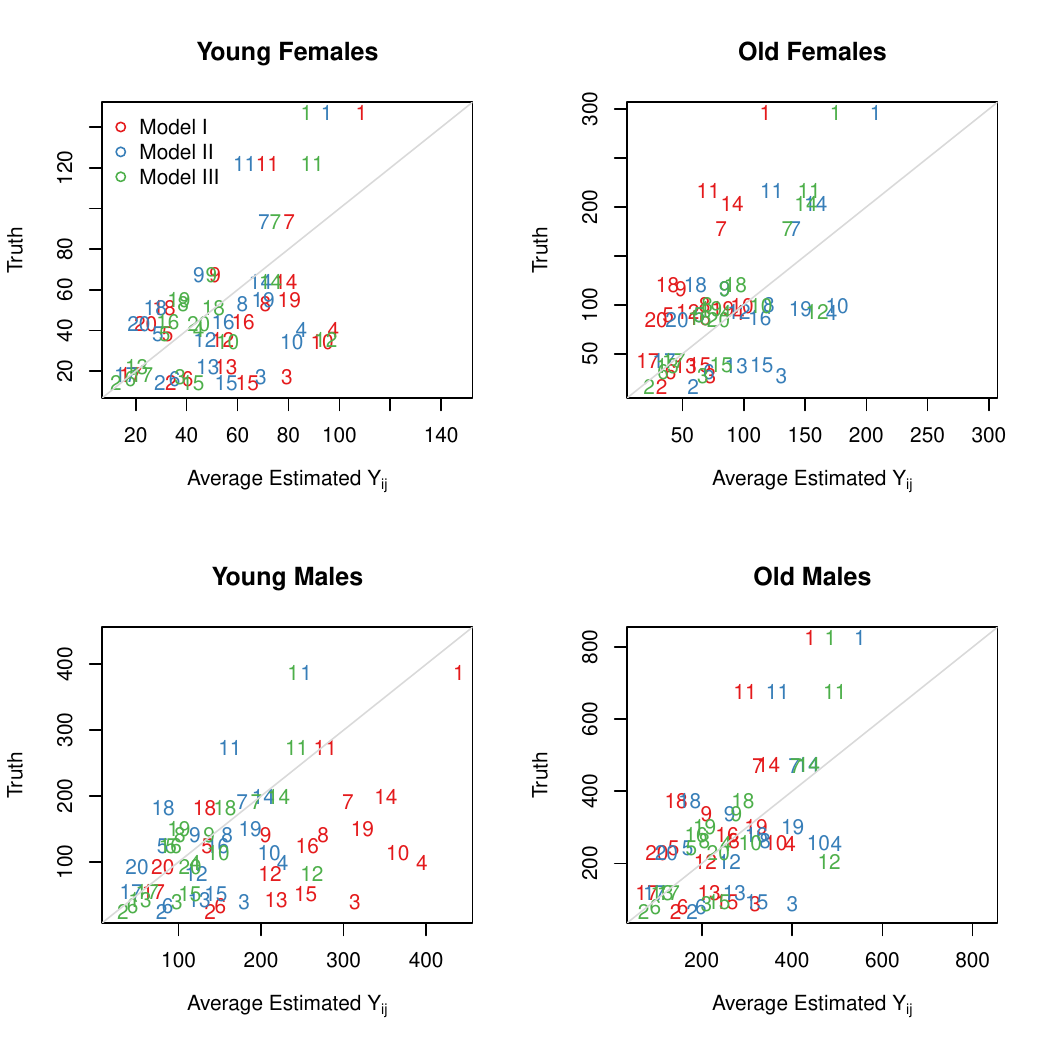}
\caption{The average village- and strata-specific estimates for the (unobserved) population counts of death plotted against the true values across each of the four models under the two-stage cluster sampling scheme for $n=5,200$. Plotting symbols indicate village numbers.}
\label{fig:Cluster_Ests_vs_Truth_n5200}
\end{figure}

\begin{figure}[htbp]
\centering
\includegraphics[width=5in]{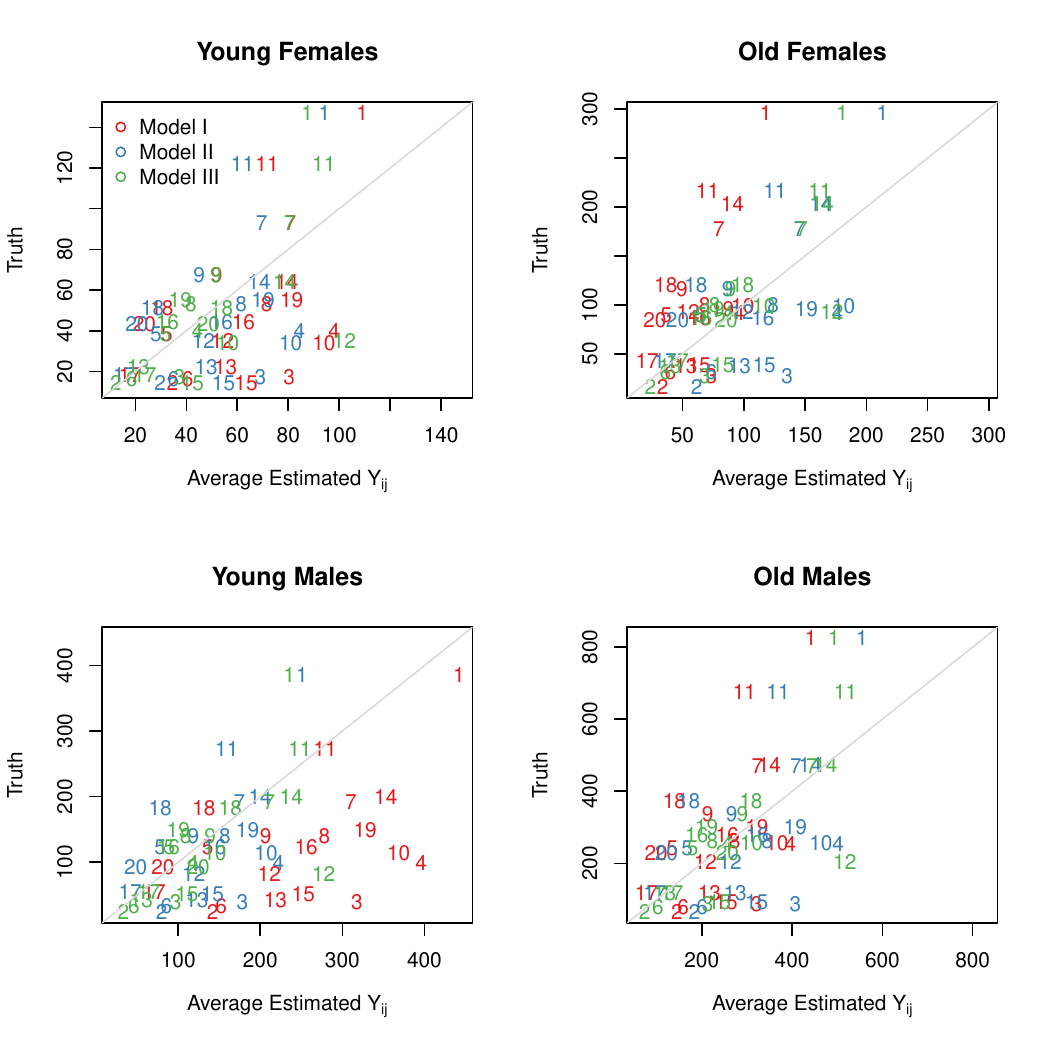}
\caption{The average village- and strata-specific estimates for the (unobserved) population counts of death plotted against the true values across each of the four models under the two-stage cluster sampling scheme for $n=3,900$. Plotting symbols indicate village numbers.}
\label{fig:Cluster_Ests_vs_Truth_n3900}
\end{figure}

\begin{figure}[htbp]
\centering
\includegraphics[width=5in]{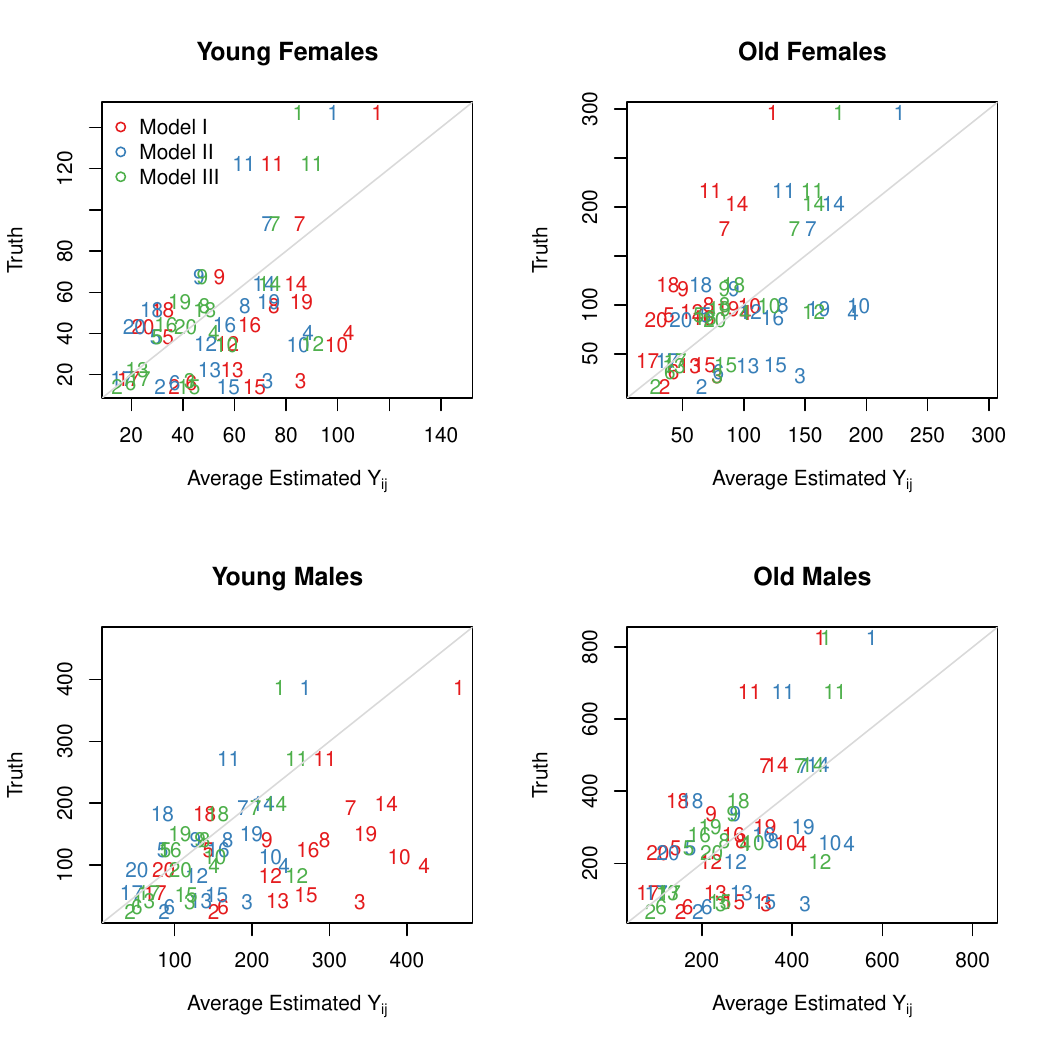}
\caption{The average village- and strata-specific estimates for the (unobserved) population counts of death plotted against the true values across each of the four models under the two-stage cluster sampling scheme for $n=2,600$. Plotting symbols indicate village numbers.}
\label{fig:Cluster_Ests_vs_Truth_n2600}
\end{figure}

\begin{figure}[htbp]
\centering
\includegraphics[width=5in]{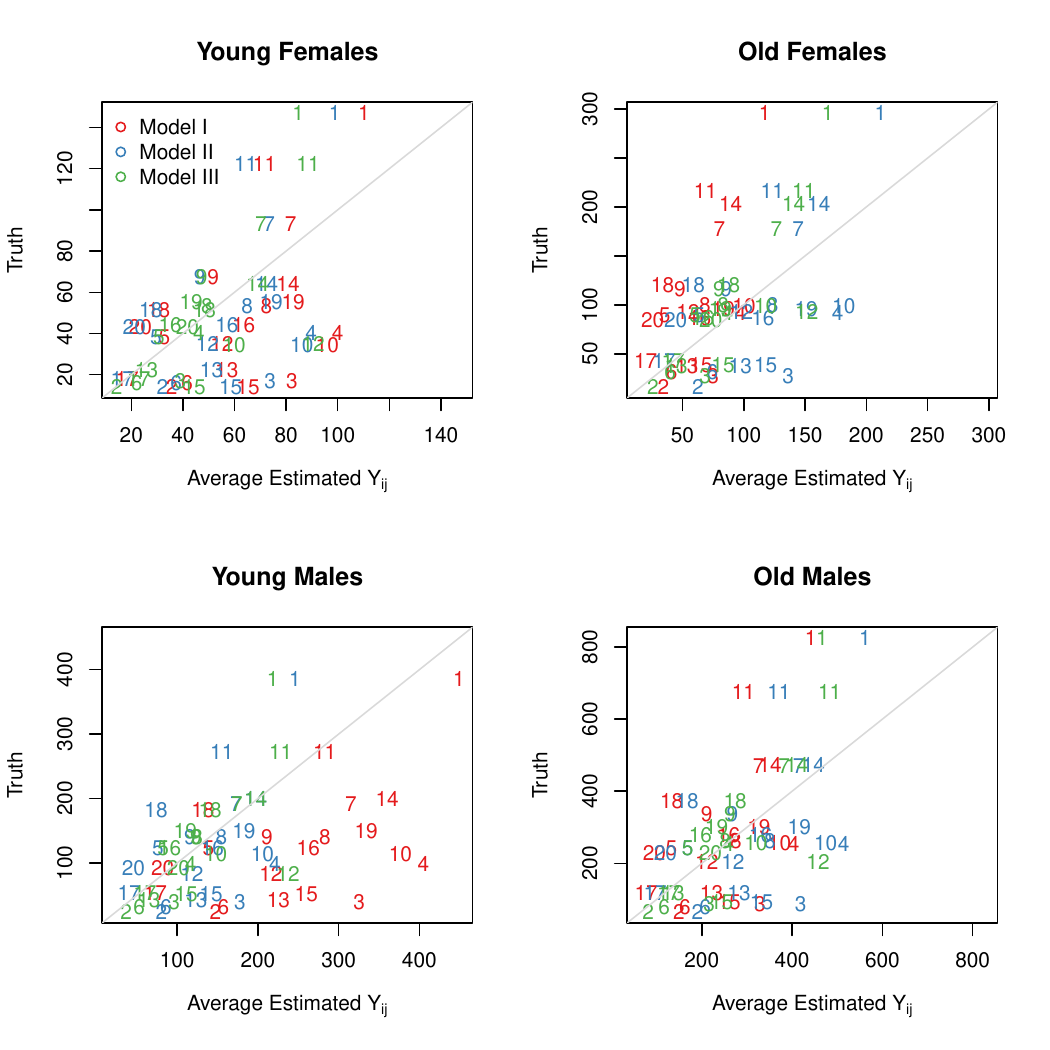}
\caption{The average village- and strata-specific estimates for the (unobserved) population counts of death plotted against the true values across each of the four models under the two-stage cluster sampling scheme for $n=1,300$. Plotting symbols indicate village numbers.}
\label{fig:Cluster_Ests_vs_Truth_n1300}
\end{figure}

\clearpage
Figures \ref{fig:Cluster_Ests_vs_Truth_n5200}-\ref{fig:Cluster_Ests_vs_Truth_n1300} display the average village- and strata-specific estimates for the (unobserved) population counts of death plotted against the true values across each of the four models under the two-stage cluster sampling scheme for $n=5,200, n=3,900, n=2,600$ and $n=1,300$, respectively.

\begin{figure}[htbp]
\centering
\includegraphics[width=5in]{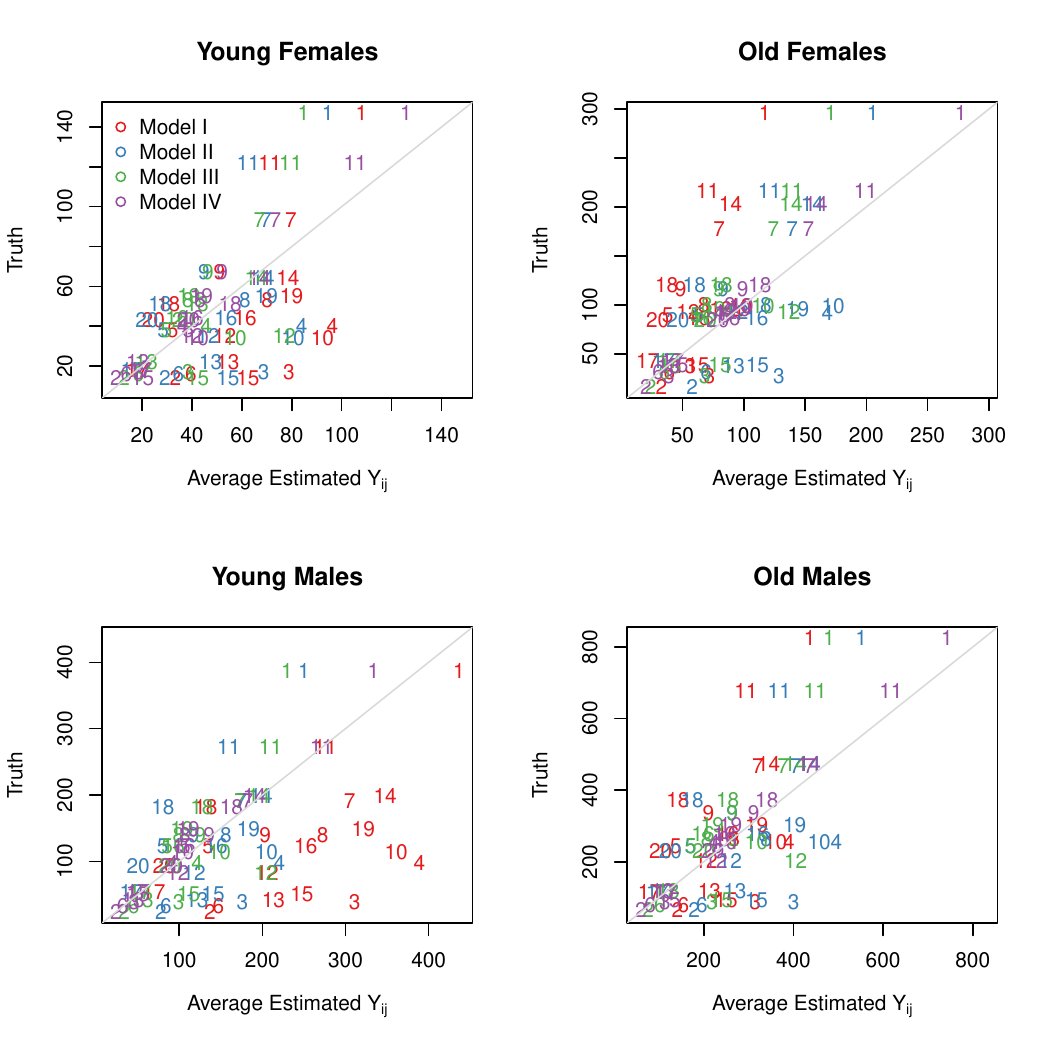}
\caption{The average village- and strata-specific estimates for the (unobserved) population counts of death plotted against the true values across each of the four models under the simple random sampling scheme for $n=5,200$. Plotting symbols indicate village numbers.}
\label{fig:SRS_Ests_vs_Truth_n5200}
\end{figure}

\begin{figure}[htbp]
\centering
\includegraphics[width=5in]{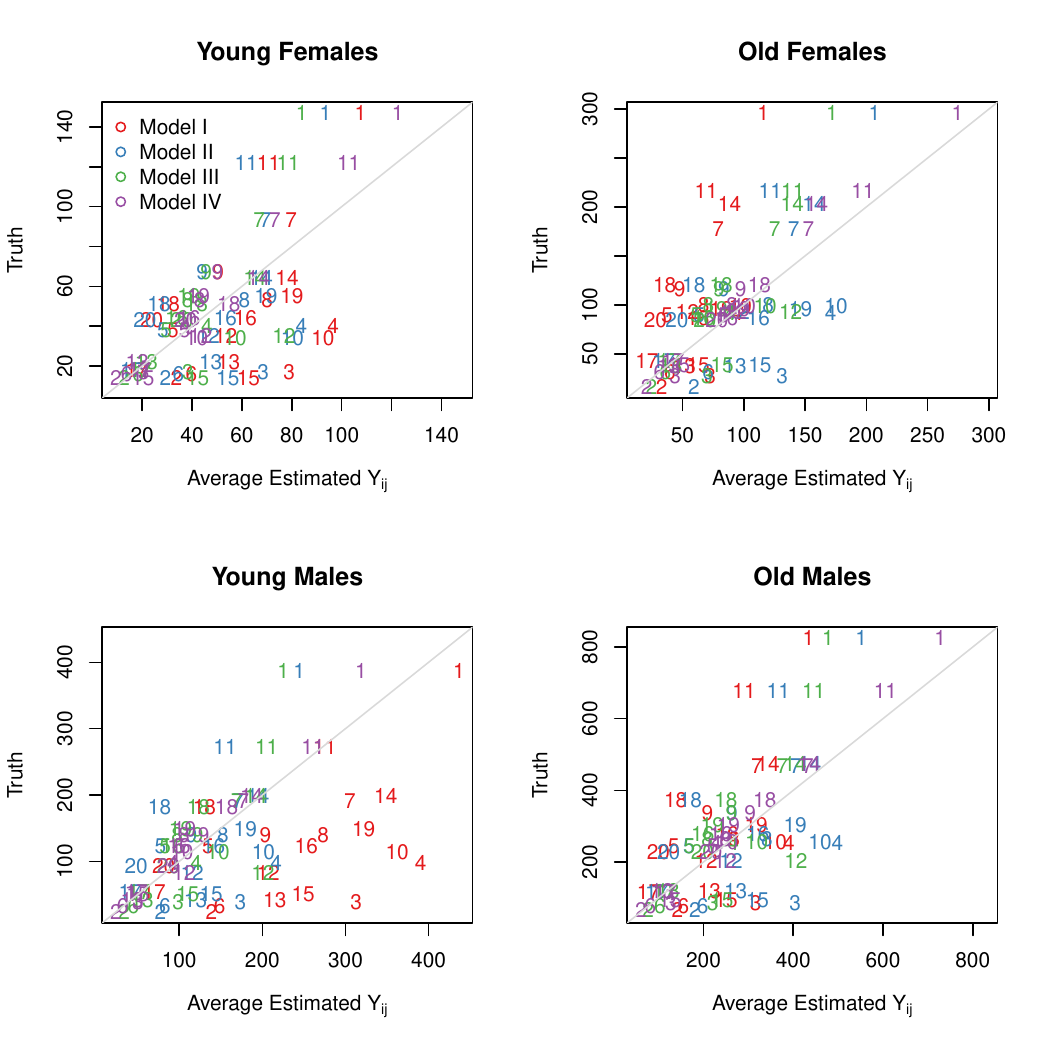}
\caption{The average village- and strata-specific estimates for the (unobserved) population counts of death plotted against the true values across each of the four models under the simple random sampling scheme for $n=3,900$. Plotting symbols indicate village numbers.}
\label{fig:SRS_Ests_vs_Truth_n3900}
\end{figure}

\begin{figure}[htbp]
\centering
\includegraphics[width=5in]{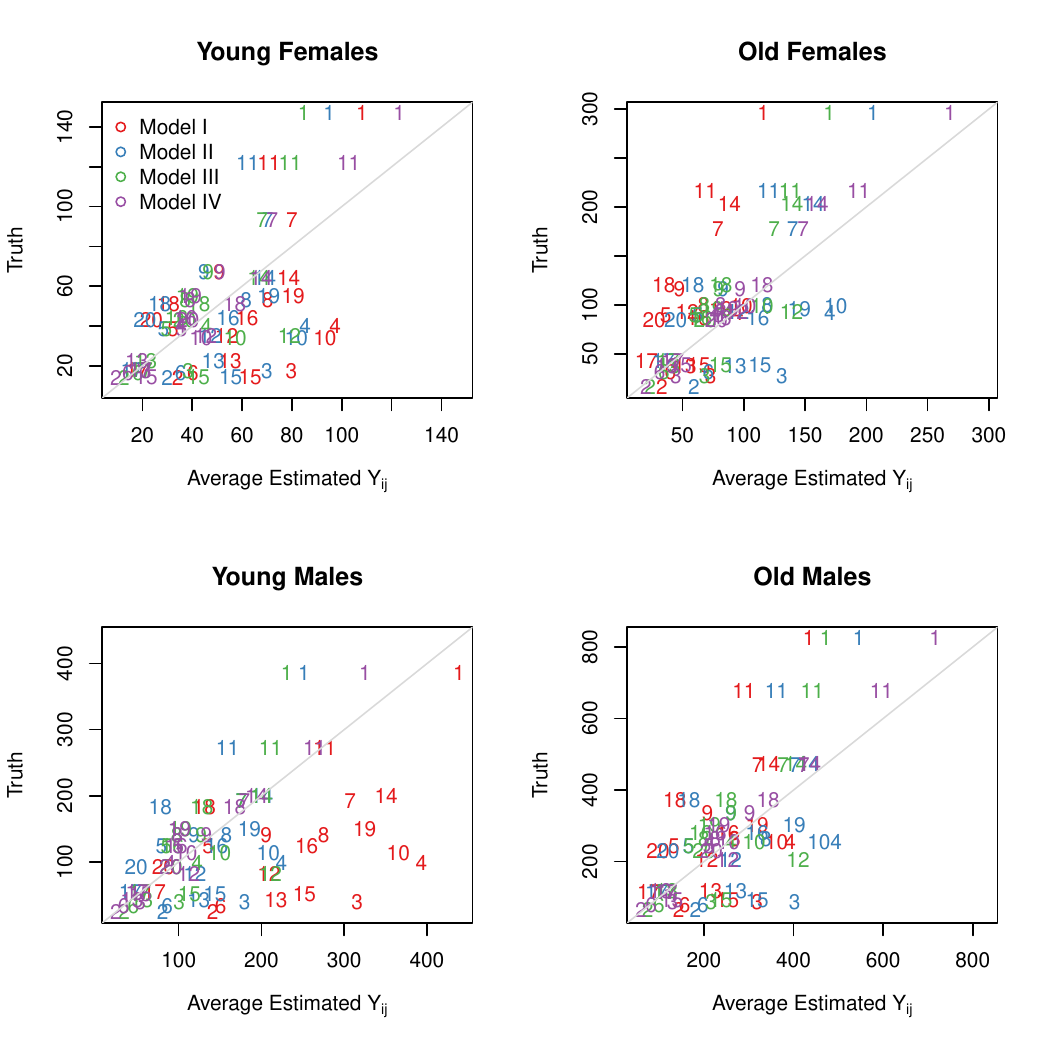}
\caption{The average village- and strata-specific estimates for the (unobserved) population counts of death plotted against the true values across each of the four models under the simple random sampling scheme for $n=2,600$. Plotting symbols indicate village numbers.}
\label{fig:SRS_Ests_vs_Truth_n2600}
\end{figure}

\begin{figure}[htbp]
\centering
\includegraphics[width=5in]{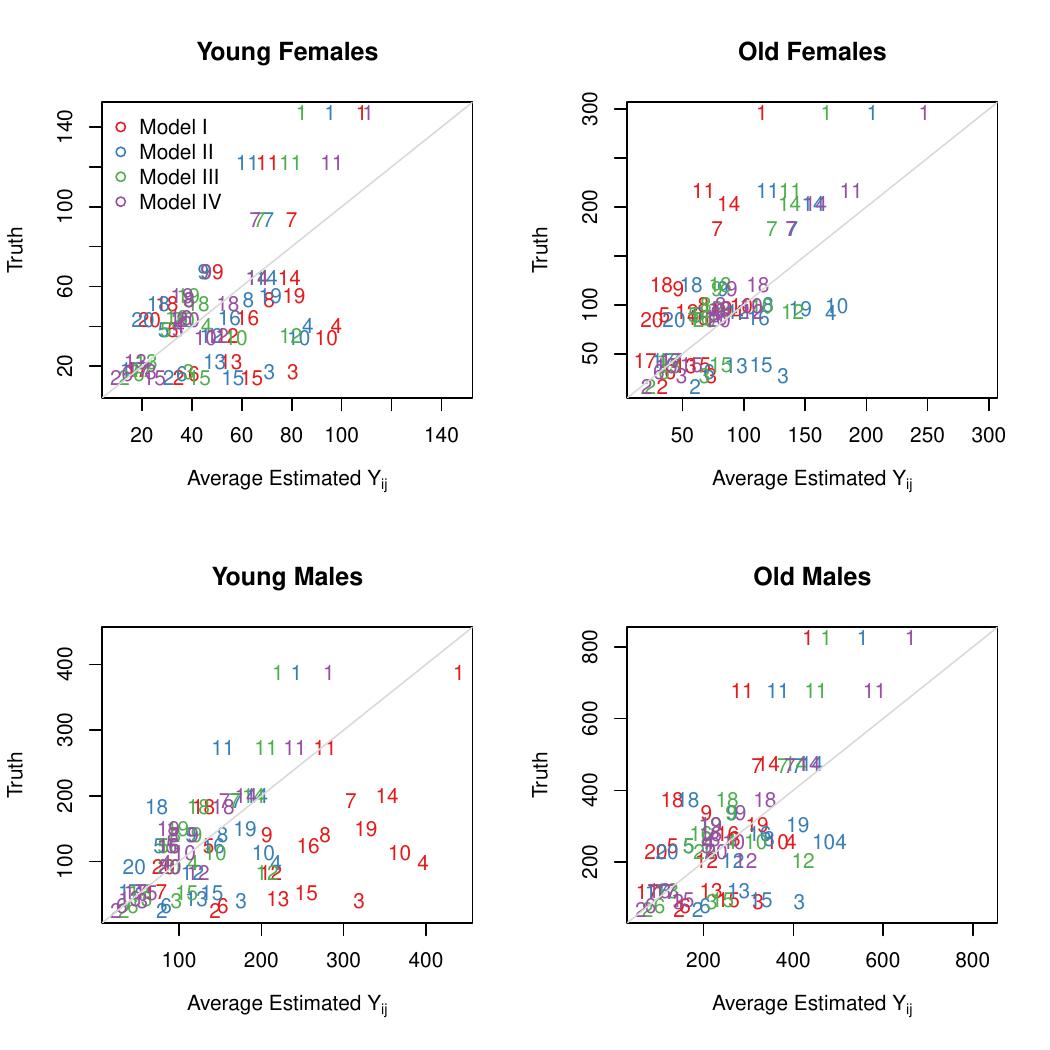}
\caption{The average village- and strata-specific estimates for the (unobserved) population counts of death plotted against the true values across each of the four models under the simple random sampling scheme for $n=1,300$. Plotting symbols indicate village numbers.}
\label{fig:SRS_Ests_vs_Truth_n1300}
\end{figure}

\clearpage
Figures \ref{fig:SRS_Ests_vs_Truth_n5200}-\ref{fig:SRS_Ests_vs_Truth_n1300} display the average village- and strata-specific estimates for the (unobserved) population counts of death plotted against the true values across each of the four models under the simple random sampling scheme for $n=5,200, n=3,900, n=2,600$ and $n=1,300$, respectively.

\begin{figure}[htbp]
\centering
\includegraphics[width=5in]{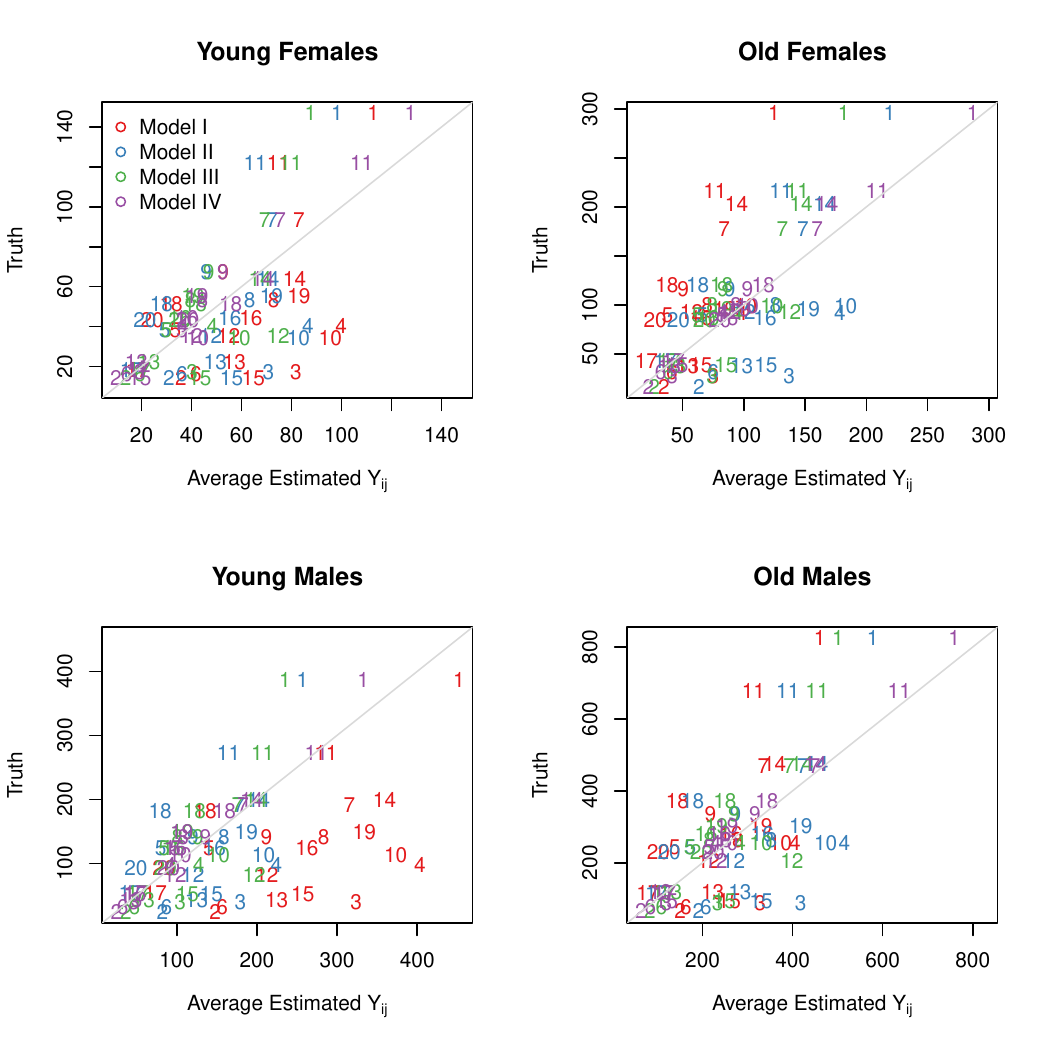}
\caption{The average village- and strata-specific estimates for the (unobserved) population counts of death plotted against the true values across each of the four models under the optimum sampling scheme for $n=5,200$. Plotting symbols indicate village numbers.}
\label{fig:Optim_Ests_vs_Truth_n5200}
\end{figure}

\begin{figure}[htbp]
\centering
\includegraphics[width=5in]{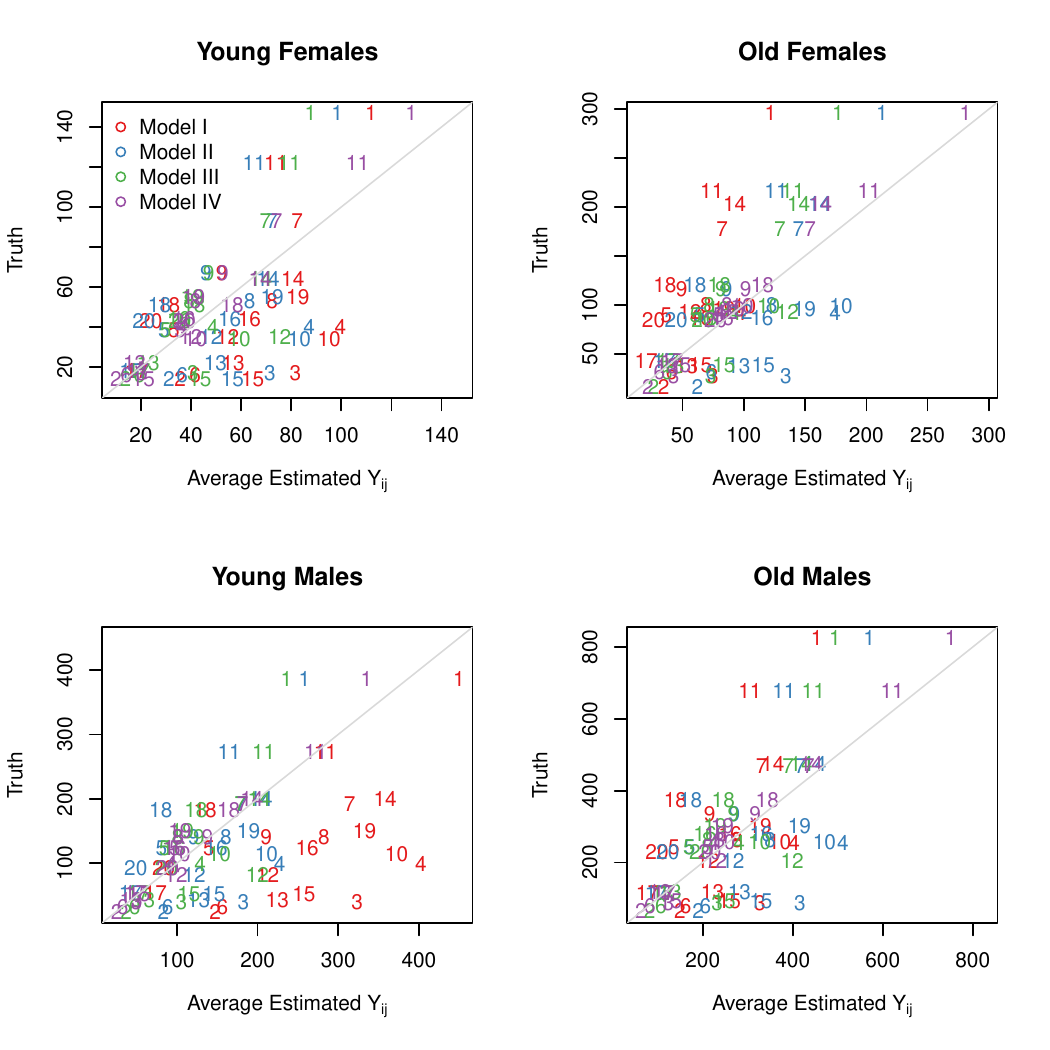}
\caption{The average village- and strata-specific estimates for the (unobserved) population counts of death plotted against the true values across each of the four models under the optimum sampling scheme for $n=3,900$. Plotting symbols indicate village numbers.}
\label{fig:Optim_Ests_vs_Truth_n3900}
\end{figure}

\begin{figure}[htbp]
\centering
\includegraphics[width=5in]{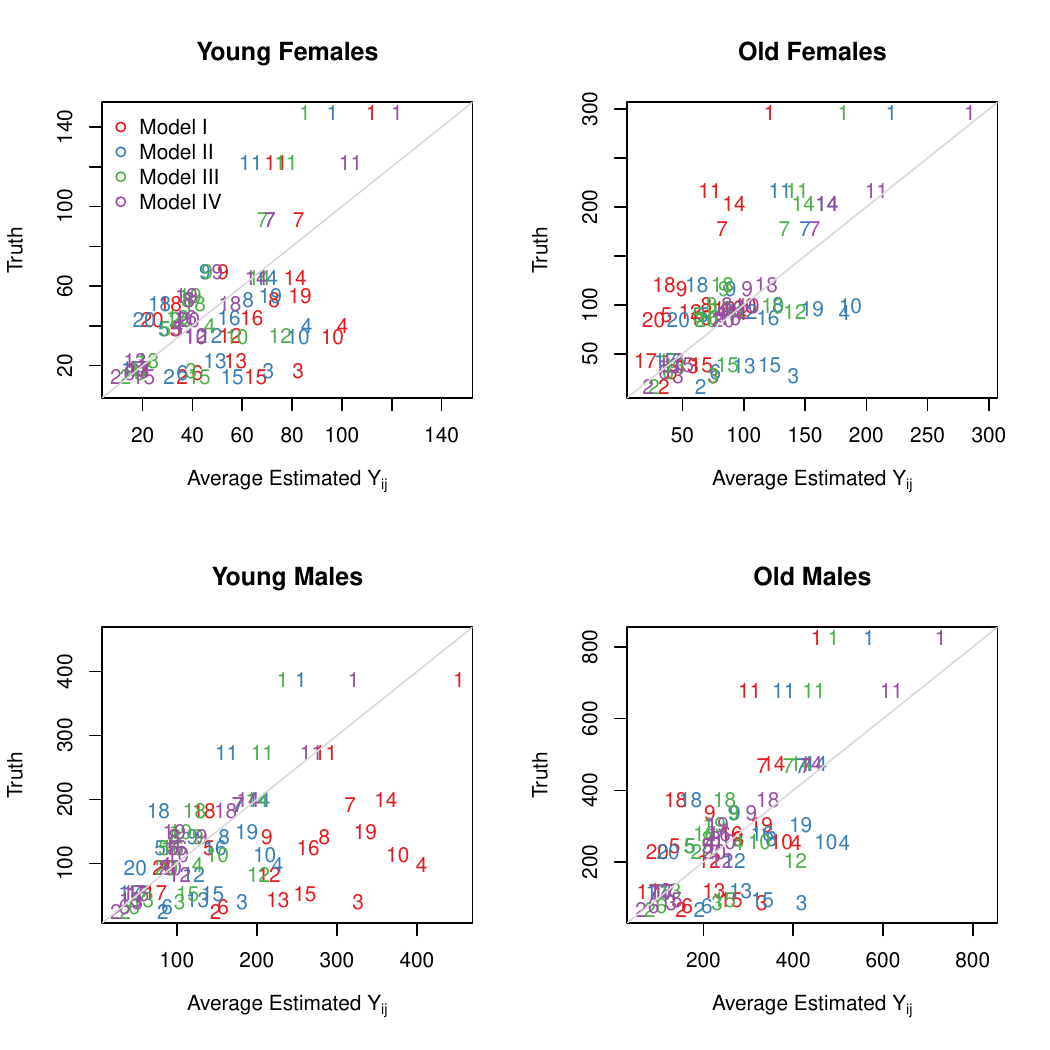}
\caption{The average village- and strata-specific estimates for the (unobserved) population counts of death plotted against the true values across each of the four models under the optimum sampling scheme for $n=2,600$. Plotting symbols indicate village numbers.}
\label{fig:Optim_Ests_vs_Truth_n2600}
\end{figure}

\begin{figure}[htbp]
\centering
\includegraphics[width=5in]{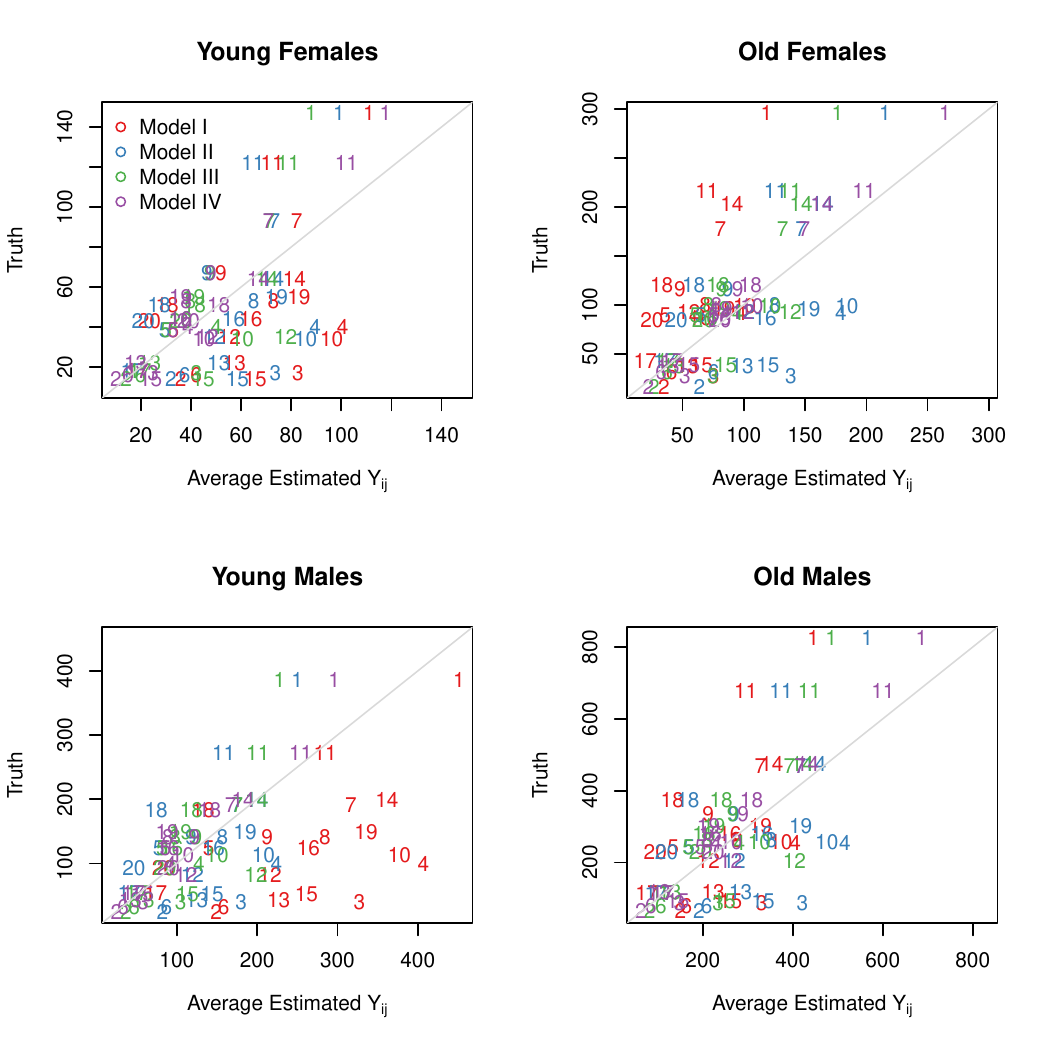}
\caption{The average village- and strata-specific estimates for the (unobserved) population counts of death plotted against the true values across each of the four models under the optimum sampling scheme for $n=1,300$. Plotting symbols indicate village numbers.}
\label{fig:Optim_Ests_vs_Truth_n1300}
\end{figure}

\clearpage
Figures \ref{fig:Optim_Ests_vs_Truth_n5200}-\ref{fig:Optim_Ests_vs_Truth_n1300} display the average village- and strata-specific estimates for the (unobserved) population counts of death plotted against the true values across each of the four models under the optimum sampling scheme for $n=5,200, n=3,900, n=2,600$ and $n=1,300$, respectively.

\end{document}